\documentclass[superscriptaddress]{revtex4-2}

\usepackage{xcolor, fancyvrb, multirow,enumitem,longtable,float}

\usepackage{hyperref}
\hypersetup{
    colorlinks=true,
    linkcolor=cyan,
    filecolor=magenta,      
    urlcolor=blue,
    pdfpagemode=UseNone
    }

\newcommand{\tablelistmargin}{10pt}

\begin{document}

\title{Quantum Undergraduate Education and Scientific Training} 
\author{Justin K. Perron}
\author{Charles DeLeone}
\affiliation{Department of Physics, California State University San Marcos}
\author{Shahed Sharif}
\affiliation{Department of Mathematics, California State University San Marcos}
\author{Tom Carter}
\affiliation{Department of Computer Science, California State University Stanislaus}
%\author{Levent Ertaul}
%\affiliation{Department of Computer Science, California State University East Bay}
\author{Joshua M. Grossman}
\affiliation{Department of Physics, St. Mary's College of Maryland}
\author{Gina Passante}
\affiliation{Department of Physics, California State University Fullerton}
\author{Joshua Sack}
\affiliation{Department of Mathematics and Statistics, California State University Long Beach}

\maketitle

%\tableofcontents

%This is the skeleton of the white paper. Hopefully I've managed to capture what we agreed upon in the meeting today in the sections as listed below. Following a suggestion from the meeting I've made two LaTeX commands \verb+\opinion{text}+ which will render the \verb+text+ in \opinion{blue} so we can highlight parts of the text that may be more our thoughts than what came from the group discussions in our breakout rooms. 

%It may be useful/easier to organize if we create separate .tex files that are used as inputs for each section.  I'll leave that up to each of you to decide what you'd like to do with your sections.

\section{Introduction}
%Quantum information science as an industry is in its infancy. The field involves disparate disciplines. 
The best way to prepare undergraduates for careers in this field is still an open question. In 2016 the United States' federal government identified quantum workforce development as a priority, citing that ``academic and industry representatives identify discipline-specific education as insufficient for continued progress in quantum information science.''~\cite{2016NSTC} More recently, the National Quantum Initiative Act was passed~\cite{NQI}, which included the goal of creating a stronger workforce pipeline; more specifically, the goal is to ``expand the number of researchers, educators, and students with training in quantum information science and technology to develop a workforce pipeline.''~\cite{NQI-housereport} Government investments like this, as well as recent significant investments from industry in quantum information science, highlight the need for a workforce with the skills and agility to thrive in this growing field.

Currently, education and workforce training in quantum information science and technology (QIST) exists primarily at the graduate and postdoctoral levels,~\cite{Passante21} with few undergraduate efforts beginning to grow out of these. In order to meet the anticipated quantum workforce needs and to ensure that the workforce is demographically representative and inclusive to all communities, the United States must expand these efforts at the undergraduate level beyond what is occurring at larger PhD granting institutions and incorporate quantum information science into the curriculum at the nation's predominantly undergraduate institutions (PUIs). On June 3rd and 4th, 2021 the \emph{Quantum Undergraduate Education and Scientific Training} (QUEST)  workshop was held virtually with the goal of bringing together faculty from PUIs to learn the state of undergraduate QIST education, identify challenges associated with implementing QIST curriculum at PUIs, and to develop strategies and solutions to deal with these challenges. The first day of the workshop included three panels:
\begin{enumerate}
	\item \underline{Establishing industry's anticipated needs}

			This panel included Heather J. Lewandowski from the University of Colorado--Boulder and JILA, staff research scientist Daniel Sank from Google, and lead quantum technologist Isabella Bello Martinez from Booz Allen Hamilton.  The goal of the panel was to inform workshop participants of what knowledge and skills are valued in the quantum industry.
	\item \underline{Learning from existing programs}

			This panel included  Emily Edwards who is the managing director of the Illinois Quantum Information Science and Technology Center at the University of Urbana-Champaign, Brian La~Cour from the Center for Quantum Research at the University of Texas at Austin, Lincoln D. Carr from the Quantum Engineering Program and Physics Department at the Colorado School of Mines, and Ezekiel Johnston-Halperin from The Ohio State University.  The goal of this panel was to inform workshop attendees about some of the approaches, successes, and challenges from existing undergraduate QIST education efforts.

	\item \underline{Fostering thriving and inclusive programs}

			This panel included Jorge A. Lopez from the University of Texas at El Paso, Robert Hilborn of the American Association of Physics Teachers, Arlene Modeste Knowles from the American Institute of Physics and project manager of TEAM-UP Diversity, and Chris Rasmussen of San Diego State University and the National Academies roundtable on Systemic Change in Undergraduate STEM Education. This panel aimed to inform workshop attendees about best practices for building thriving and inclusive programs.
\end{enumerate}

Following the panels at the end of day one, workshop attendees participated in a collaborative session aimed at identifying challenges they would face when attempting to implement what was discussed in the panels at their home institutions. The issues identified became the focus of the second day of the workshop which consisted of breakout discussions among attendees. During the breakouts attendees discussed what they felt were the key challenges they faced and aimed to develop concrete ways to approach them. This manuscript is intended to summarize the results of those breakout discussions. It is meant to assist faculty in incorporating QIST into curriculum. We discuss potential programmatic approaches to adding QIST to undergraduate curriculum in \S{}\ref{sec:ProgrammaticApproaches}, how to gain institutional support in \S{}\ref{sec:InstitutionalSupport}, curricular and content development at both the introductory (\S{}\ref{sec:FreshQIS}) and upper-division  (\S{}\ref{sec:upper-division-qist}) levels, how to incorporate inclusive practices in \S{}\ref{sec:inclusive} as well as how to a build community of support for these efforts in \S{}\ref{sec:community}. Although the details of how best to incorporate QIST into the undergraduate curriculum will be unique to each institution, we feel that these discussions and this document will help faculty address many common challenges. Furthermore, we highlight areas where additional study and resources would be very useful to the undergraduate QIST community.

%%%%%%%%%%%%%%%%%%%%%%%%%%%%%%%%%%%%%%%%%%%%%%%%%%%%%%%%%%%%%%%%

%%% Local Variables: 
%%% mode: latex
%%% TeX-master: "QUEST_whitepaper"
%%% End: 

\section{Programmatic Approaches} \label{sec:ProgrammaticApproaches}
Programmatic approaches to QIST education take a variety of forms, depending on the goals of the program, the details of the institution, and other factors.  In designing an undergraduate program, each institution must consider how to balance preparing students for immediate employment at graduation and preparing students for further study.  An institution must also consider the extent to which it prioritizes addressing immediate workforce needs versus positioning itself for future workforce demands and student interests.  Other factors include the institution’s size and the strengths/weaknesses of its departments in disciplines that contribute to QIST. 

\subsection{Types of programs}

Below, we list a variety of solutions for providing QIST education in PUIs.  The items are roughly in order of increasing scale of effort. We divide the list into approaches that likely do not need approval from a state-level or regional higher education regulatory body and those that don't.  Appendix \ref{appendix:existing} lists some examples of existing programs (compiled summer 2021).

\bigskip
\noindent \underline{\textbf{Approaches that do not likely need accreditation}}
\begin{enumerate}
  \setlength{\itemsep}{0pt}
  \item Modules in an existing course or courses
  \item New course(s)
  \item Summer institute
  \item Minor
\end{enumerate}

\noindent \underline{\textbf{Approaches that likely require accreditation}}
\begin{enumerate}
 \setcounter{enumi}{4}  
  \item Certificate
  \item Concentration or track in a major
  \item Major
  \item Master’s program
\end{enumerate}

Many faculty already include modules on QIST in existing courses.  They may accomplish such inclusion either informally by deciding on their own to add material, or formally by collective decision of the department and subsequent modification of the course description in the catalog.  A modular approach to QIST curriculum allows solutions to be built at various scales.  Such an approach is being taken by the NSF-funded Quantum Science, Technology, Engineering, Arts, and Mathematics (QuSTEAM) project of The Ohio State University, Chicago State University, University of Chicago, University of Illinois, Michigan State University, and Argonne National Laboratory with many other participants ~\cite{QuSTEAM}.  A dedicated QIST course or courses may be offered once experimentally or become part of permanent offerings.  Section \ref{sec:courses} explores these options in more depth.

A summer institute can provide another venue for trying out new curricular materials without the same level of curricular review required for courses or programs.  Summer institutes are intensive learning experiences that are often residential, though not always.  They may be offered for credit or not.  They may focus on student learning, faculty learning, or both.  Participants are not necessarily from the same institution.  Faculty and departments may find it easier to collaborate and pool resources across institutions for a summer institute than they would for courses or programs within the academic year.  A summer institute could also dovetail nicely with a summer research experience for students. (See Appendix \ref{appendix:existing} for examples.)

Creating and maintaining a program requires more effort and more resources than adding material to a course or offering a standalone course.  As compared to including QIST skills in courses, a QIST program adds additional value by providing a concrete mechanism for communicating about those skills.  It signals to prospective students that QIST training is present, and it signals to employers and graduate schools that students have those skills. 

Minors or certificates may provide means to establish a program without disrupting existing disciplinary majors. A recently published roadmap on building an undergraduate quantum engineering program~\cite{OSAroadmap} reviewed three existing minors and believed, depending on the existing majors at an institution, a quantum engineering minor can require as few as three new courses. Institutions creating a QIST minor should consider whether students should be able to complete the minor with the requisite skills drawn from various disciplinary courses (physics, computer science, math, engineering, etc.) or whether it requires at least one dedicated QIST course.  For example, the quantum engineering minor at the Colorado School of Mines may be completed without taking a QIST-specific course.~\cite{OSAroadmap}  Minors may straddle traditional disciplinary departments---a situation which may present both opportunities and challenges, as discussed below in sections \ref{sec:ProgramConsiderations} and \ref{sec:InstitutionalSupport}. 

\begin{table}[b]
\begin{tabular}{| l | c | c | }
\hline
\textbf{Approach}& \textbf{Advantages \& Opportunities} & \textbf{Disadvantages \& Challenges} \\
\hline
\multirow{3}{0.15\textwidth}{Modules}& 
\multirow{3}{0.425\textwidth}{
\begin{itemize}[noitemsep,nolistsep,leftmargin=\tablelistmargin]
\item{Little/no insitutional overhead}
\item{Useful first step to build interest and support}
\end{itemize}} 
& \multirow{3}{0.425\textwidth}{
\begin{itemize}[noitemsep,nolistsep,leftmargin=\tablelistmargin]
\item{Must replace another topic (what gets dropped?)}
\item{Student experience not formally documented}
\end{itemize}}\\
& & \\
& & \\
\hline
\multirow{4}{0.15\textwidth}{Course(s)}& \multirow{1}{0.425\textwidth}{
\begin{itemize}[noitemsep,nolistsep,leftmargin=\tablelistmargin]
\item{More room for broader coverage}
\item{Formally documented on transcript}
\end{itemize}} 
& \multirow{1}{0.425\textwidth}{
\begin{itemize}[noitemsep,nolistsep,leftmargin=\tablelistmargin]
\item{Finding room in full majors}
\item{Need sufficient interest to offer course regularly}
\item{Aligning prerequisites across majors}
\end{itemize}}\\
& & \\
& & \\
& & \\
\hline
\multirow{5}{0.15\textwidth}{Summer Institute}& \multirow{1}{0.425\textwidth}{
\begin{itemize}[noitemsep,nolistsep,leftmargin=\tablelistmargin]
\item{Intensive deep coverage}
\item{Formally recognized}
\item{Students from multiple institutions}
\item{Share/pool expertise across institutions}
\end{itemize}} 
& \multirow{1}{0.425\textwidth}{
\begin{itemize}[noitemsep,nolistsep,leftmargin=\tablelistmargin]
\item{Financial support}
\item{Logistics}
\item{Inter-institutional challenges/politics}
\end{itemize}}\\
& & \\
& & \\
& & \\
& & \\
\hline
\multirow{4}{0.15\textwidth}{Minor}& \multirow{1}{0.425\textwidth}{
\begin{itemize}[noitemsep,nolistsep,leftmargin=\tablelistmargin]
\item{Broad topic coverage}
\item{Formally recognized}
\item{Requisite skills drawn from existing majors}
\end{itemize}} 
& \multirow{1}{0.425\textwidth}{
\begin{itemize}[noitemsep,nolistsep,leftmargin=\tablelistmargin]
\item{Need sufficient faculty with relevant expertise}
\item{Significant bureaucracy/approval process}
\item{Dept.\ location/interdepartmental politics}
\end{itemize}}\\
& & \\
& & \\
& & \\
\hline
\multirow{4}{0.15\textwidth}{Certificate}& \multirow{1}{0.425\textwidth}{
\begin{itemize}[noitemsep,nolistsep,leftmargin=\tablelistmargin]
\item{Post-baccalaureate opportunity to upskill}
\item{Formally recognized}
\end{itemize}} 
& \multirow{1}{0.425\textwidth}{
\begin{itemize}[noitemsep,nolistsep,leftmargin=\tablelistmargin]
\item{Need sufficient faculty with relevant expertise}
\item{Significant bureaucracy/approval process}
\item{Dept.\ location/interdepartmental politics}
\end{itemize}}\\
& & \\
& & \\
& & \\
\hline
\multirow{4}{0.15\textwidth}{Concentration or track within a major}& 
\multirow{1}{0.425\textwidth}{
\begin{itemize}[noitemsep,nolistsep,leftmargin=\tablelistmargin]
\item{Broad topic coverage}
\item{Multiple courses incorporating QIST}
\item{Formally recognized}
\end{itemize}} 
& \multirow{1}{0.425\textwidth}{
\begin{itemize}[noitemsep,nolistsep,leftmargin=\tablelistmargin]
\item{Need sufficient faculty with relevant expertise}
\item{Significant bureaucracy/approval process}
\item{Dept.\ location/interdepartmental politics}
\end{itemize}}\\
& & \\
& & \\
& & \\
\hline
\multirow{5}{0.15\textwidth}{Major}& \multirow{1}{0.425\textwidth}{
\begin{itemize}[noitemsep,nolistsep,leftmargin=\tablelistmargin]
\item{Broad topic coverage}
\item{Multiple courses incorporating QIST}
\item{Formally recognized}
\end{itemize}} 
& \multirow{1}{0.425\textwidth}{
\begin{itemize}[noitemsep,nolistsep,leftmargin=\tablelistmargin]
\item{Need sufficient faculty with relevant expertise}
\item{Dept.\ location and interdepartmental politics}
\item{Significant bureaucracy/approval process}
\item{May be overkill for current job landscape}
\end{itemize}}\\
& & \\
& & \\
& & \\
& & \\
\hline
\end{tabular}
\caption{Summary of approaches to incorporating QIST in the undergraduate curriculum \label{tab:approach}}
\end{table}

Certificate programs have become popular with industrial employers, as they provide an efficient means to acquire a specific set of skills.  They are targeted, standalone sets of courses, and smaller than majors.  In some fields, certificates are popular at community colleges as they do not require completing a full bachelor's degree.  In other fields, certificate programs frequently make money for institutions, as they often draw working, post-baccalaureate students who seek to upskill.  However, certificate programs often need approval from state or regional higher-education accreditors.

Concentrations or tracks within traditional disciplinary majors provide another way to offer multiple courses to build QIST skills.  Institutions considering these routes should consider how many courses make a concentration or track, and what is the irreducible set of topics or courses that should be included.  For example, a program might choose to include disciplinary courses for base skills, include QIST topics within other existing disciplinary courses, and finish with a single capstone course on quantum computing (both hardware and software).  Another program might instead opt to include two upper-level courses---one on quantum algorithms and coding, and one on quantum engineering (possibly with a lab).  Concentrations and tracks likely need approval from state or regional higher-education accrediting bodies.

As of this writing, we know of no standalone QIST majors within the United States.  Industrial panelists at the QUEST workshop stated that they did not yet see the need for a standalone QIST major. This opinion is reinforced by a recent qualitative study of the quantum industry~\cite{fox2020preparing} which argued the quantum industry has a large need for employees with expertise in traditional fields but with a \emph{quantum awareness} of how their expertise affects the other elements involved in that quantum technology.  These skills and quantum awareness can be acquired within existing majors supplemented with the curricular approaches described above. Furthermore, as pointed out in~\cite{fox2020preparing}, since the quantum industry is so young the definition of \emph{quantum engineer} remains unclear. Thus, it is premature to develop a full undergraduate major at this point. In the future, the situation may change as the sector develops. There are a few institutions who have developed QIST master's programs or even BS/MS programs as described in appendix~\ref{appendix:existing}.

Table~\ref{tab:approach} summarizes the programmatic approaches described as well as some of their associated benefits and challenges.

\subsection{Program considerations} \label{sec:ProgramConsiderations}

As institutions assess plans for QIST programs, they need to pay attention to demand signals.  At the national level, industry has plenty of demand for hires, as evidenced by the National Quantum Initiative~\cite{NQI}, interviews with quantum industry representatives~\cite{fox2020preparing}, and listings~\cite{QEDCjobs} from the Quantum Economic Development Consortium (QED-C), and more.  The QUEST workshop demonstrated widespread faculty interest.  Faculty and industry need to help university administrations recognize this demand.  Student interest in QIST is increasing, but faculty and industry need to work to grow it still more.

One possible approach for creating QIST programs at PUIs is to start small and build from there.  This provides an opportunity to build student demand from existing offerings and to demonstrate that demand to university administration.  One could envision starting with an individual faculty member, then growing to multi-faculty intra-departmental activity, and finally establishing inter-departmental programming, as follows.
\begin{enumerate}
  \item Develop individual faculty expertise.
  \item Develop course modules or a course (upper-level or first-year interest course).
  \item Expand faculty interest.  Develop administration buy-in and industry connections.  Build student interest.
  \item Develop a concentration or track within a major.
  \item Develop an interdisciplinary minor or certificate.
  \item Develop a standalone QIST major (sometime in the future).
\end{enumerate}
Besides building support from demonstrated interest, this approach also allows for an exit before committing too much in case the programming does not work out. 

There are several challenges that are unique, or at least more common, at PUIs. PUIs are typically smaller than doctoral institutions which raises the issue of reaching critical mass both in terms of QIST faculty and students.  Many PUIs do not have the range of faculty expertise necessary to span the breadth of topics involved in QIST. One approach to this challenge is for PUI faculty to band together with faculty at other institutions for training in faculty summer institutes or ``Quantum Camps''. On the student side of things, to reach and maintain a critical mass it is necessary to recruit them early. First-year courses are useful for recruiting, exciting, and retaining students; however, they are difficult to fit in already crowded course sequences. This may be especially challenging at liberal arts colleges. On the other hand, upper-division courses are good for pulling together material and developing more advanced topics with students who have more prerequisite knowledge but they may come too late for many students.

Programs that span more than one disciplinary department have their own challenges.  (See also section \ref{sec:InstitutionalSupport}.)  The departments involved will need to agree on the curriculum.  They will have to figure out how to source funding for faculty training and how to compensate mentoring of interdisciplinary students.  They will also have to figure out how to divide teaching load.  With a decentralized program, it may be challenging to get buy-in from all of the needed departments, and it may be hard then to maintain it.  It can help to have administration buy-in and guidance to coordinate between departments.  Student demand is important for motivating administrators, and industry can add weight.  However, interested faculty need to coordinate to administrators the message of student demand, industry need, and external funding.  We can look to other interdisciplinary fields to see how they have successfully established and maintained programs.  In STEM, such fields include bioinformatics, nanotechnology, and materials science.  Interdisciplinary programs are even more common in the humanities.

Programs that are housed within a single department also face different challenges.  They will need to find sufficient faculty expertise within the department and possibly resources for training additional faculty.  They will also need to find capacity within the departmental faculty teaching load.  A department should also consider whether there is minimum number of faculty that it needs involved to support the program.  Beware of structuring an entire program around a single faculty member. These challenges and strategies for navigating them are discussed further in section~\ref{sec:InstitutionalSupport}.

\subsection{Topics \& skills\label{sec:skills}}
In their analysis Fox, Zwickl, and Lewandowski ~\cite{fox2020preparing} compiled a list of skills and knowledge sought by industrial employers.  That list covers mostly graduate-level hirees.  We have extracted undergraduate-level skills and topics for PUIs from it, and we have added prerequisite elements to our list that are implied by it but were not explicitly mentioned.  %The forthcoming paper by Carr \textit{et al}.\ ~\cite{} will also discuss QIST skills. 

\bigskip
\noindent \underline{\textbf{Topics}}
\begin{enumerate}  
  \item \textbf{Ideal qubits}. This topic includes idealized two-level systems, gate operations on them, and entanglement.
  \item \textbf{Physical qubits}. Students should be introduced to real physical examples such as transmons, trapped ions, and spins in semiconductors, though it may be difficult to go into very much depth at the undergraduate level.
  \item \textbf{Quantum algorithms}.  This topic includes the basic theory and more detailed implementation of algorithms and communication protocols, such as teleportation, Deutsch's algorithm, Shor's algorithm, Grover's algorithm, etc.
  \item \textbf{Noise sources}.  Include the causes of decoherence, how they are introduced, and their impacts.
\end{enumerate}

\noindent \underline{\textbf{Skills}}
\begin{enumerate}
 \setcounter{enumi}{4}
 \item  \textbf{Coding}.  This skill can be included in many courses (not only QIST-specific courses).  Include good, collaborative practices.  In many physics programs, these skills could be incorporated earlier than is currently done.  (For physics, see PICUP in Appendix \ref{app:community}.)
  \item \textbf{Statistics \& data analysis}. This skill can be included across multiple courses, including labs and including courses that are not specific to QIST.
  \item \textbf{Troubleshooting (debugging)}.  This skill can be included in lab courses, electronics courses, and programming courses --- QIST-specific and otherwise.
  \item \textbf{Modeling}.  Include learning how to increase the complexity of models to depart from ideal cases.  This skill often relies on numerical calculations and not just analytic solutions.
  \item \textbf{Electronics}.  Include circuit design, circuit characterization, and software control of hardware (e.g., LabView, Arduino, etc.).
  \item \textbf{Conducting research \& addressing open questions}. Practicing these skills early on also helps in retaining students and providing authentic experience.  Later in the college career, opportunities can be more involved (with more benefit to the research programs of faculty mentors).  
\end{enumerate}

With respect to the last item, a variety of strategies exist for incorporating research into a curriculum.  Not all experiences have to be research experiences in the narrow, traditional sense, e.g., standalone experiences or lasting all summer.  They can be ``research-like” experiences within courses.  Some programs work with clusters of students who take the same courses and then work as a group on a research project.  Putting upper-level students together with more junior students can provide impactful peer instruction and mentorship.  Even outside of classroom courses, students can get credit for research during the semester, but at some institutions it does not count toward faculty teaching load, presenting an obstacle for teaching-intensive faculty.  Many institutions are not far from large or small national labs which offer research opportunities during the summer and during the academic year.  Exploring exchanges between institutions to provide students with varied research opportunities can be fruitful.  Shared or virtual summer institutes might be a mechanism to incorporate multiple institutions.

\section{Institutional support} \label{sec:InstitutionalSupport}
Support for these efforts is often needed from multiple constituencies and at multiple levels, ranging from the departmental level to the multi-institutional level. Each level has its own intricacies and nuances. Nearly all require some amount of political acumen. Broadly speaking, within the confines of one’s particular job limitations, the motto is generally to ``just do it.'' At the departmental level, this means potentially using any free slots in your teaching schedule to teach a QIST class, or perhaps developing some kind of programming outside of the classroom that does not require approval. The idea is to demonstrate to departmental administrators that QIST can be popular and that it is worthy of support. One of the keys, especially at this departmental level, is to attempt to put systems and structures in place to support your efforts in the long run since systems and structures tend to outlast people if they are well-implemented.

\paragraph{Administrators.}

At the administrative level, as long as the administration does not have to do much and you can demonstrate a program’s benefit and worth, the administration are usually amenable. Once they see a successful program, they are more likely to support it.

That being said, we do recommend the development of a PowerPoint presentation or “glossy” PDF with employment figures and other data that can be supplied to administrators to demonstrate that QIST is a very important long-term investment and that it represents the future. This document should be developed closely with industry. In addition, if small pilot programs have been implemented in individual departments around the country, data should be taken that can be used to supplement such a document at the individual department level. In other words, localized data should be used to supplement a more broadly applicable general document. Development of such a document was identified as a priority for QUEST participants. It would be immediately useful in helping gain institutional support lowering a key barrier to incorporating QIST at PUIs.

\paragraph{Faculty and students.}

Since QIST is inherently interdisciplinary, there will often be a need to work across multiple departments. This can be tricky given the penchant for turf wars, especially at smaller institutions where enrollment numbers tend to be important markers for assessment by the administration. Addressing this challenge will be very institution specific. That being said, try to build consensus. There are models for fostering interdisciplinary work that exist. The key is creating buy-in from colleagues in multiple departments and attempting to develop a structure that works for your institution.

One possible way to inspire campus-level buy-in by students, faculty across multiple departments, and even administrators, is to create a simple, short module that introduces a topic at an accessible level that emphasizes the importance of the field. For example, one area that consistently garners interest is cryptography which has become a popular subject as the digital age matures. A good model for a very basic module that could be used to convey some basic ideas might be constructed based on ideas developed in the first three chapters of the book \emph{Protecting Information: From Classical Error Correction to Quantum Cryptography}, by Susan Loepp and William Wootters~\cite{Loepp} which is built around the broad concept of cryptography. It is a topic of broad interest to many people and provides some simple talking points for administrators.

Garnering institutional support is always easier if industry is involved. Finding support and building partnerships with industry can be facilitated through local and regional technology alliances or even simply emailing relevant people at local companies. The key — and this is universally true — is that everyone needs to benefit in some way. If industry can benefit from their involvement, obtaining their support should not be too problematic. But the College administration would likely need to see a benefit as well if they are going to support a program. Again, the document described above would facilitate demonstrating the benefit of any QIST program to administration.

\paragraph{External partners.}

One way to accomplish many of these goals is to develop cross-institutional partnerships. A consortium of several colleges is sometimes stronger than a single college, especially if there can be some sharing of resources. Approaches to developing partnerships such as these are discussed further in section~\ref{sec:community}.

\paragraph{Funding sources.}

Finally, seed funding for projects would be extremely helpful since most of the outlined ideas here require a bit of time and effort which is easier to do with some initial financial support. In addition to traditional funding mechanisms available for education efforts there is a growing number of organizations funding efforts specifically in QIST education at various scales. For example, the NSF and the Army Research Office have both communicated funding opportunities relating to QIST and education~\cite{DCL21033, LQC}, while on a smaller scale there are opportunities like the microgrant program through the UnitaryFund~\cite{UnitaryFund}. 
\newline

%{\bf Resources}\newline
%
%%How to talk to the higher-ups:
%Association of American Colleges \& Universities STEM Leadership Institute:
%
%{\hskip 1cm}{\href{https://www.aacu.org/summerinstitutes/sli/2021}{https://www.aacu.org/summerinstitutes/sli/2021}}
%
%Examples of local tech alliances:
%
%{\hskip 1cm}{\href{https://nhtechalliance.org/}{https://nhtechalliance.org/}}
%
%Example of program with local schools:
%
%{\hskip 1cm}{\href{https://www.anselm.edu/meelia-center-community-engagementcommunity-engaged-learning/access-academy}{https://www.anselm.edu/meelia-center-community-engagementcommunity-engaged-learning/access-academy}}
%
%
%

\section{Content and course design} \label{sec:courses}
\subsection{Freshmen and Sophomore QIST} \label{sec:FreshQIS}
It will be beneficial to introduce students to the QIST field as early as possible. Early introduction will promote interest in the field and enable students to make any adjustments to their academic path to suit their newfound interest. Furthermore, a strong early conceptual introduction to quantum science could help reduce the hesitancy that can arise from the ``nobody understands quantum'' narrative that is often expressed in pop culture. Participants in this breakout session felt that the most effective introduction to QIST at the lower-division level would be achieved through a new course offering. The resulting discussions were focused on exploring appropriate learning outcomes and content, necessary prerequisites, delivery options, and identifying existing resources.

A short list of concepts suitable for an introductory course in QIST at this level was created during our discussions. These align well with the ``Key Concepts for Future Quantum Information Science Learners''~\cite{KeyConcepts}. By the end of the course students should be able to:
\begin{itemize}[noitemsep]
	% \item describe, qualitatively and quantitatively, the concepts of superposition, entanglement, and quantum measurement.
	\item describe, qualitatively and formally, the concepts of superposition, entanglement, and quantum measurement
	\item list and describe examples of physical objects that can act as qubits
        \item translate between Dirac notation and list notation for vectors
	\item understand basic relevant probability and statistics
	\item implement gates and projections as matrix operators
	% \item perform the mathematical operations to calculate quantum operations such as gate transformations, changes of basis, and projections.
	%\begin{itemize}[noitemsep,nolistsep]
	%	\item for example they should know how to multiply two matrices, multiply a matrix by a vector, take the inner product of two matrices, and take a matrix or vector written in one basis and write it in terms of another basis.
	%\end{itemize}
	%\item Apply a quantum gate to a given quantum state (gate transformations)
	\item work with the basic mathematical formalism required to describe multiple qubit systems, including Kronecker products%, more to see what this means and understand how to work with them. Use the properties without getting into too much detail of the mathematics.
	%\item understand that quantum algorithms are good at taking advantage of superposition/entanglement for parallelism in the computation, but then reading off a single global property of the function or data set being studied.
	\item explain why quantum information processing has potential advantages over classical models. % \emph{i.e.}, how superposition and entanglement enable parallelism for efficient computation but that measurement and state collapse restrict the situations where this can be useful
\end{itemize}

%\comment{This section should definitely have a citation to the quantum engineering roadmap shared by Lincoln Carr and the OSA/NSF effort. Perhpaps even replace a lot of this with just a pointer to that paper as they have an entire section on developing a first QIS course and how to build it from a set of modules.} 
A roadmap for undergraduate quantum engineering programs~\cite{OSAroadmap} was recently published. It is an excellent resource for anyone looking to design a QIST curriculum. The paper outlines several modules that could be used in designing a freshman or sophomore course as well as outlining how those modules have been used in the design of existing courses. During our discussions we developed an outline of the content for such a course that echos some of the recommendations contained in the roadmap. 
\begin{itemize}[noitemsep] 
	\item QIST applications 
	\item Linear algebra %Mathematical operations and math representations of quantum states (linear algebra)
	\item One and two qubit systems %Superposition, entanglement, and quantum measurement
	%\item Mathematical functions
	\item Measurement and probability
	%\item Measurement in different bases
	\item Quantum cryptography, and random number generators
	\item Quantum algorithms (Deutsch's algorithm)
	\item Noise and quantum error correction
	\item Hardware and physical qubits
	\item Future outlook
	%\item \opinion{For hands-on experiences the course should incorporate some} quantum programming and quantum labs %\comment{I assume this means hands-on work with cloud quantum programing resources or similar items.}
	% \item Optics ased approach as an introduction - waves and polarization emphasis: gates and wave plates, starts with a practical example before becoming more abstract. 
\end{itemize}
%Participants felt strongly that at this level, the pre-requisites to an introductory QIST course should include only college algebra with 
A QIST course like this could be offered with the only prerequisite being college algebra and any additional math necessary being taught as part of the course. The lack of advanced pre-requisites ensures the course is open to students from multiple majors and various academic backgrounds. This accessibility enables students to engage early in their degrees and make any adjustments to their curricular plans they may think appropriate to pursue the field further. There are several introductory level texts currently available, a sample of which are listed in appendix~\ref{app:texts}. 

An introductory course like this could be delivered in several ways. However, we believe that interactive engagement is critical and highly appropriate for this type of course.  Additionally, the course content lends itself very well to assignments and activities requiring teamwork among students and could involve some project-based elements. These elements will help students learn to effectively collaborate and develop the soft skills that were discussed during the ``Establishing Industry's Anticipated Needs'' panel. Furthermore, these engaged collaborative approaches are among the effective practices for thriving and inclusive programs as discussed in section~\ref{sec:inclusive}.

Although designing a course as outlined above may seem straightforward, there are still open issues that require addressing. How do we best engage students so there is sufficient enrollment for regular course offerings? Are there professional development opportunities for faculty with limited experience in QIST to prepare them to offer a course like this? As the field tries to introduce QIST to students at earlier stages there will be a need for teachers at the high school and middle school levels. How can we recruit students planning on teaching to become involved in this field and how can we best prepare these teachers? These are all challenges that will need to be addressed. They also provide opportunity for those willing to engage and take the lead on developing solutions. Many solutions could be achieved through a community based approach as discussed in section~\ref{sec:community}.

\subsection{Upper-division QIST} \label{sec:upper-division-qist}
There are two main ways QIST can be incorporated into upper-division curricula: either as a new course or courses or as elements added to existing courses. Our conversation focused on ways to introduce students to QIST within an existing curriculum, pointing out topics that can be included in an existing course or a topics course.   This focus  reflects the lower barrier to this approach, which makes it more feasible in the near-term at PUIs.\footnote{For a thorough discussion on developing a new course focused entirely on QIST readers can refer to ``\emph{Building a Quantum Engineering Undergraduate Program}''~\cite{OSAroadmap}} The most obvious place for the addition of QIST-specific content would be within a quantum mechanics (QM) course in the physics department.  However, the group came up with several suggestions for other courses where small breadcrumbs of QIST content could be scattered throughout.  In the physics department, these courses included modern physics, advanced lab, statistical mechanics, applied optics, and/or electronics. In mathematics, it was thought that QIST could be part of the conversation in  linear algebra, number theory, abstract algebra, logic, graph theory, combinatorics, and/or probability.  In computer science, the courses discussed were programming methods, theory of computation, cybersecurity, data structures and algorithms, and numerical methods. 

\subsubsection{QIST in a Physics Quantum Mechanics Course}
Much of our conversation focused on the QM course in a physics department.  The group discussed the main progression of ideas in typical QM courses and discussed in particular two different approaches: instruction starting with wave functions and the Schr\"{o}dinger equation written in its differential form, and instruction starting with two-state systems emphasizing vector and ket representations of quantum states.  The group unanimously agreed that starting with two-state systems is the best way to teach quantum mechanics in an effort to include QIST content later in the semester, as most introductory QIST topics rely exclusively on two-state systems.  There are several ways to start a course in this way, the most popular being to begin with spin-1/2 particles and a Stern-Gerlach experiment as motivation, called the ``spins-first'' approach (see references~~\cite{McIntyre,Townsend,Sakurai} for common textbooks using this strategy) and photon polarization (as done in reference~~\cite{Beck}). Our conversation on how to include QIST into a QM course assumes the use of a ``two-state systems''-first approach.

Throughout our discussion, we considered three main questions: (1) what is the goal of including QIST content in a QM course, (2) how is it possible to fit QIST content into an already full course (i.e., what material can be cut or moved to another course), and (3) what QIST material could and should be included.  In response to the goal of including QIST content in a QM course, unlike what the learning goals might be in a full QIST course, the group discussion focused on cultivating an interest in QIST for future consideration and on the application (and reinforcement) of material already learned in a quantum mechanics class.  These goals imply that different institutions may find that different topics fit better with their particular course than others.  

Participants suggested two possible ways to open space for QIST content.  The first was to move some of the traditional quantum mechanics content into a modern physics course, which has experienced an identity crisis over the past several decades~\cite{Zollman16}.  The content suggested to be moved varied depending on the content already being covered in both modern physics and quantum mechanics at various institutions.  The introduction of spin-1/2 and quantum state vectors could be suitable for a modern physics course, with potentially a small introduction to the linear algebra concepts needed (such as vector and matrix multiplication and the idea of basis).  Once two-state systems are introduced, there is the opportunity to move quickly into more complex quantum mechanics topics in a QM course (such as the Schr\"{o}dinger equation and continuous wave function).  %This also allows for the possibility of including QIS topics at the modern physics level.  (For more ideas on content that is suitable for this level, see Section~\ref{sec:FreshQIS}.)

Other ideas on how to make room for QIST content in a QM course were to critically evaluate some of the topics that we traditionally spend significant time on.  As an example, several participants expressed that they felt the treatment of the hydrogen atom could be streamlined. Depending on the learning goals of a particular course, it might be possible to reduce the time spent solving this equation in class and focus on the more conceptual elements. Instructors in our conversation indicated that this was a topic that was most commonly taught by performing the mathematical procedures on the board while students follow along in their notes, which makes it an ideal candidate for moving to a supplemental video or text activity that can be completed outside of class time.  Ultimately the relevant question is ``are students better served by spending extensive time covering the mathematical details of things like the associated Laguerre polynomials and their relation to the Hydrogen atom, or by a more concise discussion of the mathematics of the Hydrogen atom and some additional QIST topics?'' Considering the former topic is not likely to be used directly in any industrial positions post graduation and would be revisited in most graduate study curriculum, workshop participants believed that the majority of our students are better served with the inclusion of QIST material. While we did not discuss other topics that may also fit this criteria, it is likely each instructor might be able to find elements of their courses where the in-class lecture portion can be shortened without skipping important content. 

Once space has been made in the course for QIST content, the question becomes what content could be included and where within the course it could be located.   To develop the list below, we brainstormed topics that are suitable (in terms of content) for a QM course and then voted on which we would most like to include in our courses.  The list below includes all topics that received more than one vote and is ordered with the most popular topics listed first.  Beneath each item is a list of prerequisite knowledge that would be helpful to include before covering each topic. Note that we are not suggesting that all the following content be included, but rather we are suggesting content that is likely suitable for the level of instruction and might fit well within a quantum mechanics course. Indeed, several of the faculty that have already included QIST topics into their course reported including only 2-4 of these topics. 
\begin{itemize}[noitemsep]
    \item Entanglement
        \begin{itemize}[noitemsep]
            \item Superposition (and bases)
            \item Multiple particles
            \item Measurement (both joint and conditional)
            %\item Density operators (and the difference between classical and quantum correlations)
           % \item Distinguishability
        \end{itemize}
    \item Quantum Key Distribution
        \begin{itemize}[noitemsep]
            \item Superposition
            \item Measurement
            %\item Mutually unbiased bases
            \item No cloning theorem
        \end{itemize}
    \item Teleportation (with or without the circuit diagram)
        \begin{itemize}[noitemsep]
            \item Entanglement (Bell states at a minimum)
            \item Measurement
            \item Quantum gates and circuit model (potentially not necessary)
        \end{itemize}
    \item Physical qubit systems and their Hamiltonians
        \begin{itemize}[noitemsep]
            \item Spin-1/2 and nuclear magnetic resonance
            \item Josephson junctions and superconductivity
            \item Photon polarization
            \item Atoms/ion trapping (and related AMO techniques)
            %\item (Potentially) Decoherence and noise sources
        \end{itemize}
    \item Bell's Theorem (tests of local realism)
        \begin{itemize}[noitemsep]
            \item Entanglement
            \item Measurement
            %\item Realism
            %\item Locality
            %\item Mixed states (potentially not necessary)
        \end{itemize}
    \item EPR Paradox and quantum steering
        \begin{itemize}[noitemsep]
            \item Entanglement
            \item Measurement
            \item Bell states
        \end{itemize}
    \item Bloch sphere (or Bloch ball)
        \begin{itemize}[noitemsep]
            \item General two-level systems
            %\item Pure and mixed states (potentially not necessary)
        \end{itemize}
    \item Quantum gates and the circuit model (towards describing the Deutsch-Josza algorithm)
        \begin{itemize}[noitemsep]
            \item States and operators
            %\item Unitary evolution operators (potentially not necessary??)
            \item Entanglement
            %\item Non-separable states
        \end{itemize}
    \item No Cloning Theorem
        \begin{itemize}[noitemsep]
            \item Superposition
            \item Unitary evolution
        \end{itemize}
    % \item Data compression
    %     \begin{itemize}[noitemsep]
    %         \item Shannon and von Neumann entropy
    %         \item Composite systems (typical subspaces)
    %         \item Density operators
    %     \end{itemize}
    % \item Superdense coding
    %     \begin{itemize}[noitemsep]
    %         \item Entanglement
    %         \item Secret communication (for photons)
    %     \end{itemize}
    % \item Quantum thermodynamics    
    %     \begin{itemize}[noitemsep]
    %         \item Density operators
    %         \item Conservation laws in QM
    %     \end{itemize}
    % \item Decoherence
    %     \begin{itemize}[noitemsep]
    %         \item Density operator and mixed states
    %         \item Open quantum systems (potentially not necessary)
    %     \end{itemize}
\end{itemize}

Items mentioned during brainstorming as ones suitable for the QM course, and topics that are very important in a more complete treatment of QIST, but ones that did not receive any votes were superdense coding, data compression, quantum thermodynamics, and decoherence.

\subsubsection{QIST in a Physics Advanced Laboratory Course\label{sec:advlab}}
The other course discussed in detail during the meeting was how QIST can be incorporating into existing laboratory courses.  In particular, the conversation focused on experiments that are relatively inexpensive and experiments that might already be in use that could have a QIST `twist'.

Experiments discussed during the meeting are listed below with references to potentially useful resources:
\begin{itemize}[noitemsep]
\item Quantum optics experiments~\cite{Beck,Thorn04,Carlson06,Branning09,Dederick14,Beck16,Ashby16,Galvez05,Galvez14}:
	\begin{itemize}[noitemsep]
	\item Spontaneous parametric down conversion 
	\item Quantization of photons (Grangier experiment)
	\item Single photon interference experiment (interferometer) + quantum eraser
	\item Bell's inequality/locality tests (entanglement experiments) 
	\item Decoherence, coherence length and filters
	%\item Decoherence experiments using coherence length ~\cite{Galvez05}
	\end{itemize}
% Below is a different list from Mark's that I merged together.
%\item \opinion{Decoherence: delay polarization components of photons with calcite leading to correlation but not entanglement}

\item NMR experiments ~\cite{TeachSpinNMR,TeachSpinPNMR}
\begin{itemize}[noitemsep]
\item how to polarize nuclei (thermal polarization) + some statistical mechanics
\item two-state systems in a magnetic field/Zeeman splitting
\item Spin-spin and spin-lattice relaxation, density matrix description, decoherence
\item pulsed NMR techniques, imaging (relates to hospital MRI)
\end{itemize}

\item Computer labs / virtual experiments (VQOL)~\cite{VQOL,Candela15,Galvez21}

\item `Cloud' quantum computing using available technologies~\cite{Qiskit,Cirq,Qsharp,AWS,QCdotcom} %(examples includ IBM Qiskit, Google's Quantum AI, Microsoft's Azure Quantum, D-Wave, and IonQ).

\end{itemize}
Although generally more expensive than building lab setups in-house, programs lacking the experimental expertise necessary can consider the various quantum educational lab setups commercially available\footnote{The mention of these products does not constitute an endorsement by the authors or workshop attendees but rather a brief list of products we are aware of. We anticipate the number of products in this space will increase steadily in the coming years}~\cite{Qubitekk,Spinflex,qutools}

During these discussions, participants stressed that one of the best resources available for advanced laboratory experiments is the Advanced Laboratory Physics Association (ALPhA)~\cite{Alpha}, an association formed to provide communications and interaction among advanced laboratory physics instructors. They host meetings as well as immersion experiences which are 2-3 days of hands-on work with a single advanced lab experiment for faculty to gain sufficient expertise and confidence to implement the lab themselves.

We also discussed the learning outcomes for these types of experiments. These included the overarching goal of reinforcing the fact that physics is an experimental science. This goal can relate directly to many of the skills sought by quantum employers and described in reference~\cite{fox2020preparing} as well as in section~\ref{sec:skills}. More experiment-specific outcomes included hands-on experience with optics equipment, reinforcing the conservation of momentum and energy in parametric down-conversion, the ability for individual photons to take multiple paths, and that entanglement and correlation are not the same.

% Learning outcomes:
% Overarching: physics is an experimental science
% Do single photons interfere with themselves? You have to do the experiment.
% Hands-on experience with optics equipment
% Reinforces conservation of energy & momentum in parametric down-conversion
% individual photons take multiple paths
% locality and reality are not simultaneously good quantum concepts
% entanglement and correlation are not the same

\subsubsection{QIST in other physics courses}

Other physics courses where QIST could be incorporated include the following.

\paragraph{Modern physics:} As discussed above the introduction of two-state quantum systems could be appropriate for inclusion in a modern physics course. This topic can serve as a starting point for many core QIST concepts such as superposition and entanglement. 
Many modern physics courses cover the time-independent Schr\"odinger equation. This topic could be followed by a survey of some of the physical qubit systems and their use in QIST.

\paragraph{Applied optics:}  Discussions of polarization naturally connect to two-state systems and can be a starting point for QIST material. From this point faculty could include many of the quantum optics experiments referenced in section~\ref{sec:advlab}.

\paragraph{Statistical mechanics:} When discussing entropy a connection could be made between Boltzmann entropy and Shannon entropy (see for example chapter~1 in \cite{QPSI}). Also, two-state systems are often used to introduce the connection between microstates and macrostates. There is potential to connect these discussions with multiqubit states.

\paragraph{Electronics:} Classical electronics will be an integral part of any quantum technology. They are necessary for the conversion of digital signals from a control processor into analog control signals that will be applied to perform qubit operations. They will also be used to convert analog measurement output from qubits into digital data a classical processor can use (see \cite{ProgressAndProspects} for more discussion). Challenges such as signal crosstalk become more serious for quantum technologies due to the lack of noise immunity of quantum gates. Furthermore, these challenges are complicated by the experimental requirements for qubit operation such as vacuum pressures and cryogenic temperatures. One must consider compatibility with things like vacuum feed throughs and thermal links when designing solutions. Introducing some of these challenges during an electronics course where students are building their expertise in classical electronics will help make them more quantum aware.

%\paragraph{Nuclear physics:} how about here?

\subsubsection{QIST in Mathematics Courses}

There are many points of entry into quantum theory in a mathematics curriculum. 
We describe below how quantum theory can be introduced in a number of different subjects.

\paragraph{Linear algebra:}

While lower-division linear algebra classes often leave little room for amendments, upper division or graduate linear algebra classes may have more flexibility.
With a basic understanding of matrices and linear maps, students may be ready for explanations on matrices for quantum gates, such as the Pauli and quantum Fourier transform matrices.
An algebra topics course focused on quantum theory could be an excellent setting for building linear algebra tools used for Shor's algorithm.
A text that explains quantum computing through the perspective of linear algebra is \cite{LiptonRegan2021}.

\paragraph{Number theory:}
In number theory, factoring of numbers provides an excellent opportunity to discuss algorithms, complexity, and the significance of Shor's algorithm.
Another opportunity may arise when discussing primitive roots and the discrete logarithm. Instructors can use Shor's algorithm as a black box to show how to solve the discrete logarithm problem and break various cryptosystems, such as RSA.
This provides an opportunity to explain the potential impact of quantum computation.

\paragraph{Abstract Algbera:}
The group of $n \times n$ unitary matrices furnishes a useful example of groups which are crucial to quantum computation. Indeed, one can think of a quantum program as being a single unitary matrix.
Searching for the order of an element could be an opportune time to mention Shor's algorithm.
In a discussion about finite cyclic groups, one can introduce the Fourier transform on a cyclic group, which is the key operation in Shor's algorithm.
After going over cosets, the hidden subgroup problem (HSP) could be introduced (see~\cite{algebra-problems}), including its relevance to cryptography. Note that while quantum algorithms are currently superior to classical ones in solving the HSP, none of these algorithms is polynomial-time. Thus it can be useful to mention that there is ongoing work on the HSP.

Time permitting, other algebraic structures can be introduced.  
Temperley-Lieb algebras can be described algebraically using generators and relations or diagrammatically, and have connections to quantum theory (see \cite{Abramsky2007}).
Frobenius algebras can help describe classical structure (such as copying orthonormal bases) within a quantum system; an introduction is given in \cite[\S 8.6.1]{CoeckeKissenger2017}.
Quantum groups, such as Hopf algebras, go beyond group theory, and may require even more background, perhaps even a full course.

There are a number of order-theoretic structures that connect to quantum theory, many of them are partially-ordered or lattice-ordered structures, often with an orthocomplementation.
For example, orthomodular lattices generalize the testable properties of a Hilbert space; in particular the lattice of closed linear subspaces of a Hilbert space forms an orthomodular lattice (see \cite{Kalmbach1983,Harding2007,Piron1976}).
More general effects are captured by Brouwer Zadeh posets; see \cite[\S 4]{DallaChiaraGiuntiniGreechie2004}.
Quantum dynamics can be generalized in the setting of quantales (complete lattice-ordered sets augmented with a monoid or semi-group operator); see \cite{Rosenthal1990,AbramskyVickers1993}.

\paragraph{Logic:}
{In a logic class that interprets propositional logic on Boolean algebras, it may be a natural step to consider various non-distributive ortholattices and the resulting quantum logics, such as orthologic, orthomodular quantum logic, and Hilbert quantum logic; see \cite{DallaChiaraGiuntiniGreechie2004}.
In complexity theory, Hilbert quantum logic provides an example of an $NP_{R}$-complete (Blum-Shub-Smale complete) problem; see \cite{HerrmannZiegler2016}.}
While the classical quantum logics express relationships among static testable properties of a quantum system, the Logic of Quantum Programs (LQP) provides a more dynamic approach capable of describing non-probabilistic properties of quantum programs (see \cite{Baltag2006} and  \cite{Bergfeld2017}).
% \opinion{While the classical quantum logics express relationships among static testable properties of a quantum system, the Logic of Quantum Programs (LQP) provides a more dynamic approach capable of describing non-probabilistic properties of quantum programs; see \cite{Baltag2006}.
% It would be illuminating to compare LQP to classical dynamic logics, such as Hennessy-Milner logic and propositional dynamic logic, as the languages are so similar.
% There are probabilistic variations of LQP, such as one in \cite{Bergfeld2017} which employs logic to prove basic properties of the BB84 quantum key distribution protocol, though such probabilistic quantum logics would be harder to fit into a course.}
{Putting logic into a different context, one could introduce Logical Bell Inequalities (see \cite{AbramskyHardy2012}, especially sections IA and IB).
}

\paragraph{Graph theory:} A quantum computation can be viewed as a walk on a directed graph of quantum states, where each eadge is given by either a projector or unitary operator.
% \opinion{Labelled transition systems can be discussed as interpretations of directed graphs with labelings of edges by ``actions".
% Any labelled directed graph could be interpreted this way.
% To make it quantum, we view the vertices as \emph{quantum states} and the edges would be quantum actions, labeled by their action.
% In quantum computation, it tends to suffice to focus on two types of actions: \emph{projectors} or \emph{unitary operators}.
% We can construct an infinite directed graph for such from a Hilbert space, by making the states the one-dimensional subspaces.
% Although this would require some background on Hilbert spaces, not much is needed beyond linear algebra.}
An alternative approach is to consider graphs whose vertex set consists of atoms in an \emph{ortholattice}, and whose edges are given by \emph{Sasaki projections} (see \cite[Section 5.1]{sep-qt-quantlog}) and ortholattice automorphisms. See \cite{BergfeldEtAl2015} and \cite{CoeckeSmets2004} for more details.
% \opinion{An alternative construction that is both simpler but less familiar is to involve orthocomplemented lattice-ordered sets (ortholattices), with additional properties that are satisfied by the lattice of closed linear subspaces of a Hilbert space. 
% Atomic ortholattices without additional properties generalizes the Hilbert space setting, and has the advantage of simplicity.
% View the atoms as the quantum states and make these the vertices of the graph.
% For arrows representing projectors and unitary operators, restrict to atoms, respectively, the \emph{Sasaki projections} (see \cite[Section 5.1]{sep-qt-quantlog}) and ortholattice automorphisms.
% See \cite{BergfeldEtAl2015} for details about the connection between ortholattices and transition systems involving additional properties.
% Orthomodularity is a natural choice for an additional property, and its connection to Saski projection is illuminating; see \cite{CoeckeSmets2004} and \cite[Definition 2.1]{BergfeldEtAl2015}.
% One could point out that a transition system built from an orthomodular lattice exhibits quantum properties (such as staying in the same state after repeating a question), and additional conditions imposed on the lattice can capture more quantum properties.
% }
{Graph theory courses often include algorithms (such as traveling salesman) and some complexity theory.  This could be an opportunity to alert the class about quantum complexity classes.}

\paragraph{Combinatorics:} 
A combinatorics course may include algorithms and some complexity theory.  This could be an opportunity to alert the class about quantum complexity classes.

\paragraph{Probability theory:}
While teaching probability theory, one can extend to discrete generalized probability on test spaces (see \cite{sep-qt-quantlog}).

% A test space is a set $\Omega$ of outcomes together with a collection of non-empty subsets of $\Omega$ called tests.
% A \emph{generalized probability function} is a function $f:\Omega \to [0,1]$, such that $f$ restricted to any test is a probability function.
Time permitting (for example, if this is a topics course), one can go into greater depth explaining how probabilities arise from inner product spaces, and cover Gleason's Theorem.
% For a graduate class, students with some background on basic measure theory and functional analysis may appreciate the book \cite{Gudder1988}.

%%%%%%%%%%%%%%%%%%%%%%%%%%%%%%%%%%%%%%%%%%%%%%%%%%%%%%%%%%%%%%%%

%%% Local Variables: 
%%% mode: latex
%%% TeX-master: "QUEST_whitepaper"
%%% End: 

\subsubsection{QIST in Computer Science Courses}
QIST could be introduced at various points in existing computer science curriculum. Some undergraduate programs offer full courses in quantum computation.  An example is a quantum computation course that covers matrix formulation of quantum computation and information, Simon’s Algorithm, Deutsch’s algorithm, quantum circuits, and Shor’s factoring algorithm, and whose prerequisite is a course on algorithms.  The book ~\cite{Nielsen2011} is a classic, though a bit challenging and may depend on the prerequisites. 

Below we describe how QIST can be inserted into courses that do not already emphasize quantum theory.
\paragraph{Programming methods:} A programming course may involve QIST by discussing basics of quantum programming in ~\cite{selinger_2004} and introducing a specific quantum programming language for example Quipper~\cite{quipper,quippertut}, Silq~\cite{silq}, or Q\#~\cite{Qsharp}.

\paragraph{Theory of computation:}
When introducing the Turing machine, variations can also be introduced, such as the quantum Turing machine.  Quantum complexity classes such as BQP can be introduced, as well as the fact that certain problems can be solved strictly faster on a quantum computer. For complexity in general, the complexity zoo~\cite{ComplexityZoo} gives a good overview.
When introducing labelled transition systems, consider transition systems whose states are states of a Hilbert space and whose actions are unitaries or projectors.

\paragraph{Cybersecurity:}
A course on cybersecurity might leave room for quantum key distribution systems, such as BB84 (for example \cite[\S12.6.3]{Nielsen2011}). Instructors can point out that common public key systems, such as Diffie-Hellman key exchange and RSA, can be broken by quantum computers, and highlight work in post-quantum cryptography. Finally, while symmetric key systems so far remain secure, Grover's algorithm necessitates doubling of key sizes.

\paragraph{Data structures and algorithms:} Comparisons of data representations using qubits versus bits and extended discussion of time and space complexity with some comparison between classical and quantum algorithms could be included in this course. Additionally discussion of probabilistic algorithms, such as primality testing, for example the RSA cryptography system, could also be included.

\paragraph{Numerical methods:} Faculty could discuss the difficulties in solving and simulating quantum systems and motivate the need for quantum computers. Discussions of fast Fourier transform and comparison with quantum Fourier transform would also be a way to include QIST in this course.

\section{Inclusive practices}\label{sec:inclusive}

\subsection{Current challenges}
\label{sec:current-challenges}

STEM in general, and physics and CS in particular, have problems with diversity, equity, and inclusion (DEI). Since QIST lies at the intersection, the problems could potentially be worse. There are many factors that have contributed to a lack of progress in correcting the DEI issues in these fields. For instance, some scientists in these disciplines have a professional bias against structures aimed at social cohesion, such as mentorship or diversity training, which are often considered ``fluffy''. DEI programs and practices do not have sufficient budgetary support, especially in the long-term. And in general, mentoring activities are not professionally recognized or rewarded.

On the plus side, DEI effective practices are usually also effective practices in general; that is, they help all students. As a result, faculty often find such practices easier to adopt. Also, there are ample resources to help diversify programs, such as the \href{https://www.aps.org/programs/innovation/fund/idea.cfm}{APS IDEA Network}, the \href{https://engage.aps.org/stepup/home}{STEP UP program}, and the \href{https://www.aip.org/diversity-initiatives/team-up-task-force}{TEAM-UP} project.

In the remainder of this section, we discuss how to recruit underserved populations, how to retain these populations once they've been recruited, and pedagogical measures that support diversity efforts. 

\subsection{Recruitment}
\label{sec:recruitment}

It is crucial to reach out to students early in their academic careers, including during their high school years. Accordingly, recruitment efforts should involve the larger community. Departments should cultivate and seek funding for community ties. Recruitment events can then easily include targeted efforts for middle school and high school students.  These efforts can leverage QIST's pop culture appeal through applications like cryptography, quantum teleportation, quantum money, quantum machine learning, quantum computing, and the quantum internet.

There are a variety of effective ways to reach out to students.
Hands-on experiences are particularly effective. These experiences can occur in the first year of college, but also at the high school level.  They can involve working in labs on relevant technology; for instance, students could work with polarizers, gates, or circuit board design.
Note that the last refers to \emph{classical} circuits. Hands-on experience can also be virtual, such as with the Virtual Quantum Optics Laboratory~\cite{VQOL}.

Another type of formative experiences are mentorship opportunities with researchers. These can be faculty, but there are also opportunities with professionals in research laboratories. For instance, the Department of Defense offers many opportunities for students to work alongside department researchers~\cite{DoDSTEM}. Successful graduates can also return and talk about their experiences. This provides a direct opportunity for students to see that ``people like them'' can belong and even thrive in the field.

Publicizing job databases can also make clear the social relevance of QIST education. Consider publicizing job resources such as the Quantum Computing Institute at Oak Ridge National Laboratory's mailing list~\cite{ORNLmail} and the QED-C's job postings~\cite{QEDCjobs}. Showing students that there are relevant career tracks, and that a QIST career can benefit their own communities is an effective recruitment tool. 

 % this link was dead so I added QED-C one instead. Can we find the correct Harrisburg one? \url{https://jobs.quantumapalooza.com/questEvent/}.

Finally, recruitment efforts should cast a broad net. General audience talks and other outreach (such as the Adopt-a-physicist program~\cite{PhysAdopt}) can reach a wide cross-section, including middle- and high-school students. Curriculum should be designed with multiple pathways having low barriers to entry. Once early learners have been introduced to the field their path into formal programs should be straightforward.

\subsection{Retention}
\label{sec:retention}

Although recruitment is a necessary requirement for effective inclusive practices it is not by itself sufficient. Getting a representative population into a program should not be considered successful if those students do not succeed at completing the program. Therefore, successful programs must also focus efforts and resources on retention. Fortunately, many of the recruitment practices are also good for retention. The main additional goal is to make sure students feel they are engaged in a welcoming and supportive community.

A key obstacle for undergraduates, and especially under-represented students, is financial stress~\cite[p. 43]{Teamup}. Funding student activities can help reduce this stress and have a large impact on retention.  Faculty should identify and implement student funding opportunities whenever possible. 

Early research experiences are effective recruitment tools but they are also instrumental in student retention~\cite{Spinup,Teamup,graham2013increasing}. In course-based research experiences, student teams design and perform experiments, and they analyze the data.  Students in these courses have greater ownership of their experiments, and they experience the intellectual challenges and excitement of authentic empirical practice.  Non-course based research experiences (such as REUs or capstone projects) provide additional opportunities for in-depth research experience and one-on-one mentoring, as well as financial support.  Faculty should \emph{relentlessly} pursue students to join them in early research experiences and provide funding for
them to participate.

% Consider requiring a research experience for graduation (cf. SIU--Carbondale and Loyola University in Chicago) \comment{Shahed: I could not confirm that these institutions require research}. \\ \comment{Josh G.: Without evidence, I'm not convinced of the retention value of requiring a (non-course based) research experience simply at some point before graduation.  There may be learning value, but that's something else.  There IS documentation of retention value to EARLY (often course-based) research experiences.  At SMCM, we require all physics majors to complete a research experience before graduation (capstone research project for the physics major; applied physics majors have more options: capstone, REU, internship, Directed Research courses), but it's for the learning outcomes and some general recruitment --- no evidence of impact on retention.}

Showing students the big picture in terms of career, job opportunities remains key~\cite{Teamup,EP3}. Successful programs engage with industry to ensure students and graduates are aware of job opportunities. An exemplary program is the \href{http://fresnostate.edu/engineering/vip/}{Valley Industry Partnership}~\cite{VIP} hosted by California State University Fresno. Relationships with industry should be bilateral.  Industry can tailor their job advertisements to better recruit available
talent, including populations that might not be obvious at first sight.

Industry is also a resource for mentorship opportunities, which can demonstrate to students that careers are out there for people like them. Ideally, industry mentors should provide role models with similar identities to students. But even publicizing profiles of underrepresented minorities currently working in QIST can be inspiring to students. Older students can also serve this role and provide mentoring through peer-leadership and learning assistant programs.

For faculty mentorship, it is important to have institutional support, since mentoring activities are usually not formally rewarded, and active mentors risk burning out. Furthermore, mentorship responsibilities, as well as all other DEI efforts, should be shared by faculty across a program to ensure these efforts are sustainable. The following resources for mentors, including potential grants, can be helpful:
\begin{itemize}
    \item \href{https://aps.org/programs/minorities/nmc/}{APS National Mentoring Community} \url{https://aps.org/programs/minorities/nmc/}
    \item \href{https://www.aps.org/careers/advisors/index.cfm}{APS Tools for career advisors} \url{https://www.aps.org/careers/advisors/index.cfm}
    \item \href{https://www.aps.org/careers/guidance/advisors/bestpractices/index.cfm}{Best practices for educating students about non-academic jobs} \url{https://www.aps.org/careers/guidance/advisors/bestpractices/index.cfm}
    \item \href{https://www.mathprograms.org/db}{AMS math programs} \url{https://www.mathprograms.org/db}
    \item \href{https://terrytao.wordpress.com/career-advice/}{Career advice} \url{https://terrytao.wordpress.com/career-advice/}
\end{itemize}

Other opportunities for students can be found at
\begin{itemize}
    \item \href{https://www.aps.org/programs/women/cuwip/}{Conferences for Undergraduate Women in Physics} \url{https://www.aps.org/programs/women/cuwip/} 
    \item \href{https://www.spsnational.org/about/partnerships/nsbp}{National Society of Black Physicists (NSBP) and Society of Physics Students (SPS) joint membership} \url{https://www.spsnational.org/about/partnerships/nsbp}.
\end{itemize}

Programs should also consider adopting a code of conduct for students and instructors. A code of conduct can send a message that programs are serious about inclusiveness. One example, adapted from San Francisco State University faculty member Federico Ardila~\cite{axioms}, is a set of four axiomatic beliefs:
\setlist[enumerate,1]{leftmargin=2.cm}
\begin{enumerate}[start=1, label={Axiom~\arabic*:}]
    \item Scientific potential is distributed equally among
  different groups, irrespective of geographic, demographic, and
  economic boundaries.
    \item   Everyone can have joyful,
  meaningful, and empowering experiences with science.
    \item  
  Science is a powerful, malleable tool that can be shaped and used
  differently by various communities to serve their needs.
    \item  
  Every student deserves to be treated with dignity and respect.
\end{enumerate}

\subsection{Pedagogy}
\label{sec:pedagogy}

Some pedagogical practices that help with inclusion and retention include project-based, team-based labs~\cite{Feder17}. For instance, see Heather Lewandowski's advanced lab work~\cite{Zwickl13,Wilcox16,Dounas17}, especially if it can be adapted to introductory courses.

Active learning has been shown to have a positive impact on underrepresented groups. Some research-based active learning measures include Peer-Led Team-Learning~\cite{Gosser} and the Learning Assistants (LA) Program (\url{https://learningassistantalliance.org/}). Learning assistant programs also benefit students by providing a source of income for those serving as LAs, thereby reducing the financial burden that disproportionally affects minority groups~\cite{Teamup}. Furthermore, these programs are useful for retention because they involve students in leadership positions which is also effective for increasing retention.

Due to its novelty, QIST poses some challenges for students who might be scared that they are not talented enough, or have the right background, to explore the field. Faculty should emphasize a growth mindset and consider talking openly about their own difficulties~\cite{Walton11}. Students can also be steered towards free online learning resources. These include online courses, such as the Coursera course on \href{https://www.coursera.org/projects/getting-started-quantum-machine-learning}{Quantum Machine Learning} or the EdX courses on \href{https://www.edx.org/learn/quantum-computing}{quantum computing}. There are also numerous tutorials for quantum computing platforms like Qiskit. These resources can give students the opportunity to become comfortable with QIST at their own pace and with low stakes. Alternatively, these tools can be used to get students prepared for research experiences.

One open question is: How do we promote use of effective teaching practices among faculty? Any single course can be taught by a variety of faculty using different approaches, and many lower-division courses are taught by contingent faculty. One method is to encourage participation in pedagogical workshops and learning communities, such as the Mathematical Association of America's \href{https://www.maa.org/programs-and-communities/professional-development/project-next}{Project NExT}, or the \href{https://www.aapt.org/Conferences/newfaculty/}{New Faculty Workshop} sponsored by the American Association of Physics Teachers (AAPT) the American Physical Society (APS)  and the American Astronomical Society (AAS).

\subsection{Engagement with community colleges}
\label{sec:engag-with-comm}

Community colleges serve an important role in the educational pipeline. For example, 51\% of California State University graduates start their higher education in community college~\cite{CCCreport}. Furthermore, community colleges tend to have more diverse student populations. Thus, strengthening the CC-PUI pipeline would likely have a high impact.  These efforts can span simple efforts like targeting advertising of QIST opportunities towards CC students, or more involved efforts like aligning CC and PUI curriculum to ease the transition of transfer students and minimize time to graduation. Forming inter-institutional partnerships, like the North County Higher Education Alliance~\cite{NCHEA} can be powerful tools to strengthen these pipelines. Establishing partnerships like this are discussed further in section~\ref{sec:community}.

For more formal activites, one potential funding avenue is the \href{https://www.nsf.gov/funding/pgm_summ.jsp?pims_id=5464}{NSF Advanced Technological Education (ATE) grant program}. These grants are intended to establish partnerships between 2-year institutions and industry. But also consider examples like \href{https://www.nsf.gov/awardsearch/showAward?AWD_ID=1836033}{grant 1836033}, which seeks to measure the impact of course-based research with regards to industry-specific skills. A similar effort focused on QIST courses would be valuable.

\subsection{Further reading}
\label{sec:further-reading}

\begin{itemize}
    \item \href{https://www.aps.org/programs/education/undergrad/faculty/spinup/spinup-report.cfm}{SPIN-UP report}~\cite{Spinup}. Commissioned by the APS, the SPIN-UP report studied factors which lead to greater growth in physics departments.
    \item \href{https://www.compadre.org/JTUPP/report.cfm}{Phys21: Preparing physics students for 21st century}. The Joint Task Force on Undergraduate Physics Programs authored this report, with the purpose of helping physics programs prepare their majors for careers in industry.
    \item \href{https://ep3guide.org/}{Effective Practices for Physics Programs (EP3)}~\cite{EP3}. An APS/AAPT collaboration which collects best practices in an assortment of domains for use by physics departments.
    \item \href{https://www.aip.org/diversity-initiatives/team-up-task-force}{TEAM-UP report}~\cite{Teamup}. The TEAM-UP report commissioned by the AIP studied obstacles facing African-American physics majors, and how best to surmount them. 
    \item \href{https://www.nap.edu/catalog/25568/the-science-of-effective-mentorship-in-stemm}{Effective mentoring}~\cite{NAP25568}. Put together by the National Academies of Sciences, Engineering, and Medicine, this book guides departments towards effective mentoring relationships, especially as pertains to DEI.
    \item \href{https://www.nature.com/articles/d41586-019-00039-7}{Mentoring article}~\cite{Woolston19}. A Nature article highlighting recipients of the 2018 Nature mentoring award.
\end{itemize}

%%%%%%%%%%%%%%%%%%%%%%%%%%%%%%%%%%%%%%%%%%%%%%%%%%%%%%%%%%%%%%%%

%%% Local Variables: 
%%% mode: latex
%%% TeX-master: "QUEST_whitepaper"
%%% End: 

\section{Building community} \label{sec:community}
Many of the challenges with implementing QIST at PUIs identified by workshop participants had potential solutions rooted in a community of QIST educators. Benefits of a strong community include providing support for its members through formal and informal professional development opportunities, providing a network to share resources and materials, and in some cases providing financial support. Communities can also provide advocacy and a sense of identity for its members. Although there are clear and obvious benefits associated with a community of QIST educators the details of what that community/communities would look like are less clear. Communities have been defined according to geographical region from local communities at a given institution, regionally involving multiple institutions, or even nationally. They can also be defined by the members' discipline of study, or type of institution. We envision that a broad thriving QIST education community ecosystem would be made up of a network of subcommunities serving different populations with overlapping goals and interests. We identified some specific purposes these subcommunities could serve, the potentially interested populations, and the geographical span(s) most appropriate for that purpose. These are described below and summarized in Table~\ref{tab:community}.

Curriculum development is an area where community could have a significant impact. Developing or locating existing QIST student learning outcomes, instructional materials and lab activities were all mentioned as challenges facing faculty trying incorporate QIST into their programs. A group of faculty could be formed to develop, collect, and share these materials, assess the efficacy of materials, and explore new approaches to QIST instruction. These community efforts would ease the burden on individual faculty. Furthermore, PUIs often lack certain resources involved in a strong QIST educational program. For example a given institution may lack the hardware necessary to provide certain hands on experiences, or an institution may lack the diversity of expertise among faculty to capture the breadth of the QIST field. Sharing resources within a community is an obvious way to overcome these challenges.

Interdisciplinary communities would also be very beneficial to faculty at PUIs. For example, physics faculty members would benefit from discussion and collaboration with math and computer science faculty to better capture the interdisciplinary nature of QIST. Collaborative communities formed across departments and colleges are going to be necessary to design optimal curricula as well as navigating any of the political challenges that arise with interdisciplinary efforts.
Finally, to develop a robust workforce pipeline there must be smooth onboarding into the QIST field for students arriving at PUIs as well as students transitioning into the workforce or graduate study. On the front end, inter-institutional collaboration between regional high schools, community colleges and PUIs is needed to create interest and excitement in the QIST field among early learners as well as to inform them about the opportunities available.  On the back end, collaboration between PUIs, industry partners, and graduate institutions is needed so that PUI students can be given a clear picture of the opportunities available upon graduation and how best to pursue them.

\begin{table}
\begin{tabular}{| l | l | l |}
\hline
\textbf{Primary purpose} & \textbf{Groups involved} & \textbf{Geographical span} \\
\hline
\multirow{2}{17 em}{Building interdisciplinary connections} & Departments and colleges within an academic institution, & local, regional, \\
& Different academic institutions, Industry partners. & national. \\
\hline
\multirow{2}{17 em}{Share QIST educational materials}& PUIs, community colleges, & local, regional,  \\
& High schools, Ph.D granting institutions & national. \\
\hline
\multirow{2}{17 em}{Share QIST educational resources}& PUIs, community colleges, & local, regional, \\
& Ph.D granting institutions & national. \\
\hline
\multirow{2}{17 em}{Student research experiences}& PUIs, PhD granting institutions, & local, regional, \\
& Industry partners, Government labs & national.\\
\hline
\multirow{1}{17 em}{Graduate employment and networking}& PUIs, Industry partners & local, regional, national. \\
\hline
\multirow{1}{17 em}{Outreach and recruitment}& PUIs, Community colleges, High schools& local, regional.\\
\hline
\end{tabular}
\caption{\label{tab:community}Goals of communities within a broader QIST education community as identified by workshop participants. The potentially interested groups and appropriate geographical span of the communities are also listed.}
\end{table}

The first and obvious step toward establishing a community is to decide on a focus or purpose for the community. Are you hoping to increase recruitment into your QIST program vs. for example, establish a network of faculty to create and share educational resources?  With a clear purpose for the community, what is required for that community to thrive? Workshop participants considered existing communities they considered successful (see Appendix~\ref{app:community} for some examples) and identified four elements that were common across them:
%\begin{enumerate}[noitemsep,nolistsep]
\begin{enumerate}[noitemsep]
\item Champions and leadership
\item Resources and support
\item Visibility
\item Sustained activity.
\end{enumerate}

For almost any project to be successful it has to have someone championing it. Someone with enthusiasm for the goals of the project and who can advocate and promote it across all stakeholders. Relating to community, this person would identify and recruit others with similar goals and relevant expertise and establish some leadership for the community. Leadership should match the goals of the community as well as reflect the diversity of the community. Here we mean diversity in all senses: diversity of discipline or field of study, institution type, and demographics. Leadership responsibilities will vary depending on the details of the community. However, one common responsibility is to establish a clearly defined goal, this can be through a vision or mission statement. Other responsibilities will likely include liasing between the organizations involved, identifying and acquiring support (see below), performing outreach and promoting visibility. Ultimately these responsibilities will determine the necessity for named positions and councils in the community. Formalizing these roles will be useful when applying for funding. Furthermore, named roles will enable the people filling those roles to include them in promotion and tenure files. The formation of community leadership will likely be a cyclic process. A small group of people championing the idea will establish the goals of the community and identify, where necessary, a few others to join the leadership. Input from the new members may slightly modify some of the detailed roles required resulting in recruiting additional leadership members.

Although there are several examples of communities that thrive through members volunteering their time and effort, there is always some level of support required. Financial compensation for members dedicating significant time, digital infrastructure for hosting webpages, communications, and outreach, rental of conference facilities and travel support are all potential needs of a community that will require support. Ways to support these and other needs must be identified and acquired. Many needs can be met through in-kind resources available through member institutions but often needs must be met through other avenues such as external funding or membership dues. Thanks to the National Quantum Initiative act~\cite{NQI} and other federal efforts there is an increasing number of funding avenues that could potentially support QIST education. For example, during the workshop Dr. Abiodun Ilumoka from the Division of Undergraduate Education of the NSF presented some of their programs that could support QIST related work including two dear colleague letters, \emph{Advancing Quantum Education and Workforce Development}~\cite{DCL21033} and \emph{Advancing Educational Innovations that Motivate and Prepare PreK-12 Learners for Computationally-Intensive Industries of the Future}~\cite{DCL20101}. Other federal funding agencies working in QIST can be found at \url{www.quantum.gov}. Funding from state government agencies like the California State University system, or professional societies like the American Physical Society and the Mathematical Association of America should also be considered.

In addition to strong leadership and support, successful communities tend to have good visibility and sustained activity. Outreach should be performed to make sure potentially interested members are aware the community exists. For QIST we recommend leveraging existing communities with overlapping interest to achieve this. Consider including announcements in talks at annual meetings like APS's annual March Meeting or the AAPT/AAS/APS Physics \& Astronomy New Faculty Workshop. 

Leveraging existing networks can help increase the visibility of the community but ultimately this visibility needs to be accompanied by regular activities. There should be annual events that could take the form of meetings, workshops, summer schools, or similar events where the community will interact in a productive and meaningful way. There should also be asynchronous communications such as a quarterly newsletter or active online discussion forums. Almost by definition, the hallmark of a successful community is regular productive activity. Leadership should be intentional and deliberate in how they cultivate this activity, especially early on. It must also be noted that a clearly defined and enforceable code of conduct should be in place as toxic interactions are detrimental and undermine the entire community effort.

With QIST being a relatively new field there is a need to establish a QIST education community. Establishing an organized community which includes participation of PUI faculty would help address many of the challenges discussed throughout the workshop and in this manuscript and should therefore be a priority moving forward.

\section{Summary}
Incorporating QIST into the undergraduate curriculum is a growing priority. To meet the anticipated workforce needs of the quantum industry the United States will need to involve academic institutions beyond those larger PhD granting institutions which currently house the bulk of the academic research efforts in this field. The nation's PUIs are ideally suited to reach a broad range of students. Furthermore, bringing QIST to the diverse student populations served at many PUIs as the field is still developing provides an opportunity to avoid some of the DEI issues that plague the STEM fields. However, bringing effective QIST education into the undergraduate curriculum is not an easy task. We have identified multiple complex and interrelated issues that PUI faculty will have to overcome in this effort. We have also suggested various strategies and identified resources that should be useful in addressing these issues.  How to best incorporate QIST material in the undergraduate curriculum is very institution specific; however, we believe the suggestions and resources discussed in this manuscript will be broadly applicable, and we hope they will help faculty introduce QIST concepts to their students.

\begin{acknowledgments}
%\ack
This work is funded by the American Physical Society (APS) Innovation Fund. The APS Innovation Fund provides funding to advance collaborative projects that support the APS mission ``to advance and diffuse the knowledge of physics for the benefit of humanity, promote physics, and service the broader physics community.''
\end{acknowledgments}

\appendix
\section{Existing QIS programs}
\label{appendix:existing}
Compiled in summer 2021, this list is almost certainly incomplete. We expect many other programs to be created in the near future.  Still, these examples may be useful for others looking to start their own programs.  %We recommend
\begin{enumerate}
  \item Summer institutes
  \begin{itemize}
    \item \href{https://staq.pratt.duke.edu/summer-school}{Duke University's Quantum Ideas Summer School}
    \item \href{https://uwaterloo.ca/institute-for-quantum-computing/useqip}{University of Waterloo's Undergraduate School on Experimental Quantum Information Processing}
    \item \href{https://www.lps.umd.edu/summer-of-quantum-workshop/}{University of Maryland's Summer of Quantum Workshop}
    \item \href{https://enrichment.harrisburgu.edu/qca/}{Harrisburg University's Quantum Computing Academy}
  \end{itemize}
  \item Minors
  \begin{itemize}
    \item \href{https://quantum.mines.edu}{Colorado School of Mines}
    \item \href{http://collegecatalog.uchicago.edu/thecollege/molecularengineering/#Quantum%20Information%20Science}{University of Chicago}
  \end{itemize}
  \item Certificates
  \begin{itemize}
    \item \href{https://wayfinder.utexas.edu/degrees/certificate-quantum-information-science}{University of Texas at Austin}
    \item \href{https://quantum.mines.edu}{Colorado School of Mines}
    \item \href{https://www.colorado.edu/initiative/cubit/education-workforce-training}{University of Colorado Boulder} (Professional Masters, Bachelor and certificate programs planned) 
  \end{itemize}
  \item Concentrations or tracks in disciplinary bachelor's degrees
  \begin{itemize}
    \item \href{http://collegecatalog.uchicago.edu/thecollege/molecularengineering/#summaryofrequirementsforthemajorinmolecularengineeringquantumengineeringtrack}{University of Chicago} (quantum engineering track with BS in molecular engineering) 
  \end{itemize}
  \item Standalone majors
  \begin{itemize}
    \item \textit{None known in the United States}
    %\item \href{https://www.colorado.edu/initiative/cubit/education-workforce-training}{University of Colorado Boulder} (Professional Masters, Bachelor and certificate programs planned) 
    \item \href{https://degrees.unsw.edu.au/bachelor-of-engineering-honours-quantum-engineering/}{University of New South Wales Bachelor of Quantum Engineering}
  \end{itemize}
  \item Master's programs
  \begin{itemize}
    \item \href{https://quantum.mines.edu}{Colorado School of Mines} (separate quantum hardware and software tracks)
    \item \href{https://ece.duke.edu/masters/study/quantum-computing}{Duke University} (quantum computing concentrations (software or hardware) with a research-oriented MS or an industry-focused MEng in electrical and computer engineering) 
    \item \href{https://qsec.gmu.edu/education/}{George Mason University} (concentration within the MS in Engineering and Applied Physics program)
    \item \href{https://www.harrisburgu.edu/programs/ms-next-generation-technologies/}{Harrisburg University of Science \& Technology} (QIS concentration in Next Generation Technologies Masters Degree)
    \item \href{https://www.cqse.ucla.edu/education.html}{University of California Los Angeles} (interdisciplinary program set to begin in Fall 2022) 
    \item \href{https://dornsife.usc.edu/physics/msqis/}{University of Southern California} (MS in quantum information science)
    \item \href{https://pdc.wisc.edu/degrees/quantum-computing/}{University of Wisconsin at Madison}  (MS in physics--quantum computing)
    \item \href{https://master-qe.ethz.ch/}{ETH Z\"{u}rich Master in Quantum Engineering}
    \item \href{https://uwaterloo.ca/institute-for-quantum-computing/}{University of Waterloo} (QIS specialization tracks within traditional MS and PhD)
    \end{itemize}
\end{enumerate}

%%%%%%%%%%%%%%%%%%%%%%%%%%%%%%%%%%%%%%%%%%%%%%%%%%%%%%%%%%%%%%%%

%%% Local Variables: 
%%% mode: latex
%%% TeX-master: "QUEST_whitepaper"
%%% End: 
 
\section{Introductory level textbooks\label{app:texts}}
Several textbooks are available that could serve as required or supplemental texts for an introductory course on quantum information science. During the workshop many were identified and are listed in table~\ref{tab:texts} below. This list is by no means a complete list and many more texts are likely to be published in the near future. 

\begin{table}[H]
\begin{tabular}{| l |l| l | }
\hline
\textbf{Title}& \textbf{Author} & \textbf{Publisher} \\
\hline
Quantum Physics: What Everyone Needs to Know & Michael G. Raymer & Oxford University Press \\
\hline
Q is for Quantum & Terry Rudolph & Free Press \\
\hline
Quantum Computing: an applied approach & Jack D. Hidary & Springer \\
\hline
Quantum Computing for Everyone & Chris Bernhardt & The MIT Press \\
\hline
\multirow{2}{0.4\textwidth}{Quantum Computing for the Quantum Curious}& \multirow{2}{0.35\textwidth}{Ciaran Huges, Joshua Isaacson, Anastasia Perry, Ranbel F. Sun and Jessica Turner} & \multirow{2}{0.25\textwidth}{Springer}\\
& & \\
\hline
\multirow{2}{0.4\textwidth}{Quantum Computing A Gentle Introduction}& \multirow{2}{0.35\textwidth}{Eleanor G. Rieffel and Wolfgang H. Polak} & \multirow{2}{0.25\textwidth}{The MIT Press}\\
& & \\
\hline
\multirow{3}{0.4\textwidth}{Learn Quantum Computation using Qiskit}&\multirow{3}{0.35\textwidth}{Abraham Asfaw et al.} & \multirow{3}{0.25\textwidth}{IBM digital textbook. \url{https://qiskit.org/textbook/preface.html}} \\
& & \\
& & \\
\hline
\end{tabular}
\caption{Introductory level QIST textbooks\label{tab:texts}}
\end{table}
 
\section{Examples of Successful Communities\label{app:community}}
During our discussions we identified several existing communities with some relation to QIST eduction. The communities listed in table~\ref{tab:communities} can be used as models for successful communities as well as a resource for collaboration and strengthening a QIST education community.

\begin{table}[H]
\begin{tabular}{| l | c | l | }
\hline
\textbf{Community}& \textbf{Audience} & \textbf{Details} \\
\hline
\multirow{2}{13 em}{American Physical Society (APS)}& \multirow{2}{6 em}{Physics} & \multirow{2}{30 em}{Broad community of physicists spanning all fields.}\\
& & \\
%\hline
%\multirow{2}{13 em}{APS - Division of Quantum information}& \multirow{2}{6 em}{Physics}& \multirow{2}{30 em}{A division of APS focused on QI} \\
%& & \\
\hline
\multirow{3}{13 em}{National Society of Black Physicists (NSBP)}& \multirow{3}{6 em}{Physics}& \multirow{3}{30 em}{Organization devoted to the growth, development, and advancement of the African-American physics community.} \\
& & \\
& & \\
\hline
\multirow{2}{13 em}{American Institute of Physics (AIP)}& \multirow{2}{6 em}{Physics}& \multirow{2}{30 em}{Organization with mission to advance, promote, and serve the physical sciences for the benefit of humanity.} \\
& & \\
\hline
\multirow{2}{13 em}{Society of Physics Students (SPS)}& \multirow{2}{6 em}{Physics}& \multirow{2}{30 em}{American Institute of Physics Student organization.} \\
& & \\
\hline
\multirow{2}{13 em}{American Association of Physics Teachers (AAPT)}& \multirow{2}{6 em}{Physics education}& \multirow{2}{30 em}{Organization dedicated to enhancing the understanding and appreciation of physics through teaching.} \\
& & \\
\hline
\multirow{4}{13 em}{Partnership for Integration of Computation into Undergraduate Physics (PICUP)}& \multirow{4}{6 em}{Physics education}& \multirow{4}{30 em}{Community of educators that support the development and improvement of undergraduate physics education through integration of computation across its curriculum.} \\
& & \\
& & \\
& & \\
\hline
\multirow{4}{13 em}{American Mathematical Society (AMS)}& \multirow{4}{6 em}{Mathematics}& \multirow{4}{30 em}{Professional society with aim of advancing and connecting the diverse global mathematical community through publications, meetings and conferences, MathSciNet, professional services, advocacy and awareness programs.} \\
& & \\
& & \\
& & \\
\hline
\multirow{2}{13 em}{Mathematical Association of America (MAA)}& \multirow{2}{6 em}{Mathematics}& \multirow{2}{30 em}{World's largest community of mathematicians, students, and enthusiasts.} \\
& & \\
\hline
\multirow{3}{13 em}{Society of Industrial and Applied Mathematics (SIAM)}& \multirow{3}{6 em}{Mathematics}& \multirow{3}{30 em}{International community which fosters the development of applied mathematical and computational methodologies needed in various application areas.} \\
& & \\
& & \\
\hline
\multirow{4}{13 em}{Association for Computing Machinery (ACM)}& \multirow{4}{6 em}{Computer Science}& \multirow{4}{30 em}{As the world's largest computing society, ACM strengthens the profession's collective voice through strong leadership, promotion of the highest standards, and recognition of technical excellence.} \\
& & \\
& & \\
& & \\
\hline
\multirow{3}{13 em}{Computer Science Teachers Association (CSTA)}& \multirow{3}{6 em}{Computer Science education}& \multirow{3}{30 em}{Community of computer scientists who shares best practices in K-12 CS education.} \\
& & \\
& & \\
\hline
\multirow{4}{13 em}{Society for the advancement of Chicanos/Hispanics and Native Americans in Science (SACNAS)}& \multirow{4}{6 em}{Science}& \multirow{4}{30 em}{Organization dedicated to fostering the success of Chicanos/Hispanics and Native Americans, from college students to professionals, in attaining advanced degrees, careers, and positions of leadership in STEM.} \\
& & \\
& & \\
& & \\
\hline
\multirow{2}{13 em}{American Association for the Advancement of Science}& \multirow{2}{6 em}{Science}& \multirow{2}{30 em}{Organization which seeks to advance science, engineering, and innovation throughout the world for the benefit of all people.} \\
& & \\
\hline
\multirow{2}{13 em}{Council on Undergraduate Research (CUR)}& \multirow{2}{6 em}{Science}& \multirow{2}{30 em}{Support and promotes high-quality mentored undergraduate research, scholarship, and creative inquiry.} \\
& & \\
\hline
\multirow{2}{13 em}{Institute of Electrical and Electronics Engineers (IEEE)}& \multirow{2}{6 em}{Engineering}& \multirow{2}{30 em}{The world's largest technical professional organization dedicated to advancing technology for the benefit of humanity.} \\
& & \\
\hline
\multirow{2}{13 em}{Society of Women Engineers (SWE)}& \multirow{2}{6 em}{Engineering}& \multirow{2}{30 em}{National society empowering women to achieve their full potential in careers as engineers and leaders.} \\
& & \\
\hline
\multirow{3}{13 em}{National Society of Black Engineers (NSBE)}& \multirow{3}{6 em}{Engineering}& \multirow{3}{30 em}{National society that supports and promotes the aspirations of collegiate and precollegiate students and technical professions in engineering and technology.} \\
& & \\
& & \\
\hline
\multirow{3}{13 em}{North County Higher Educational Alliance (NCHEA)}& \multirow{3}{6 em}{Regional STEM}& \multirow{3}{30 em}{A higher education consortium working to improve opportunities for North County citizens through collaboration between Cal State San Marcos, MiraCosta College, and Palomar College.} \\
& & \\
& & \\
\hline
\multirow{3}{13 em}{Michigan Academy of Science, Arts and Letters (MASAL)}& \multirow{3}{6 em}{Regional STEM}& \multirow{3}{30 em}{Regional professional association that forsters scholarly dialogue through annual conferences and quarterly journal that includes papers and news about ongoing research at Michigan institutions.} \\
& & \\
& & \\
\hline
\end{tabular}
\caption{Examples of successful communities\label{tab:communities}}
\end{table}

\section{Workshop Participants}

\begin{longtable}[H]{l c r}
%\begin{tabular}{ l  c  r }
\hline
\hline
\textbf{Name}& \textbf{Institution} & \textbf{Department} \\
\hline
\hline
Charles Adler 	& St. Mary's College of Maryland & Physics \\
Mark Beck	& Reed College & Physics \\
Mohsen Beheshti	& California State University Dominguez Hills & Computer Science \\
Sean Bentley 	&  Adelphi University & Physics \\
Sagar Bhandari 	& Slippery Rock University & Physics \\
Sambit Bhattacharya & Fayetteville State University & Computer Science \\
Henry Boateng 	& San Francisco State University & Mathematics \\
Sherrene Bogle	& Humboldt State University & Computer Science \\
Peter Brereton 	& United States Naval Academy & Physics \\
Russel Caballos	& Chicago Quantum Exchange & Quantum Education \\
Paul Cadden-Zimansky & Bard College & Physics \\
Tom Carter	& California State University Stanislaus & Computer Science \\
Albert Chan	& Fayetteville State University & Computer Science \\
Tianran Chen	& West Chester University & Physics \& Engineering  \\
Robert Chun	& San Jose State University & Computer Science \\
Lilian Clairmont& Appomattox Regional Governor's School & Physics \& Calculus \\
David Collins	& Colorado Mesa University & Physics and Environmental Sciences \\
Charles De Leone& California State University San Marcos & Physics \\
Erin De Pree 	& St. Mary's College of Maryland & Physics \\
Ian Durham	& Saint Anselm College & Physics \\
Jay Erker	& California  Polytechnic State University, San Luis Obispo & Chemistry \\
Levent Ertaul 	& California State University East Bay & Computer Science \\
Midhat Farooq	& American Physical Society	& 	\\
%Mic Fenech	& The University of Texas at Austin & STEM Education \\
Terrill Frantz	& Harrisburg University & QISC \\
Merideth Frey	& Sarah Lawrence College & Physics \\
Enrique Galvez 	& Colgate University & Physics and Astronomy \\
Katharina Gillen& California Polytechnic State University, San Luis Obispo & Physics \\
Joshua Grossman	& St. Mary's College of Maryland & Physics \\
Peng Guo	& California State University Bakersfield & Physics and Engineering \\
Katie Hall 	& Wellesley College & Physics \\
Erik Helgren	& California State University East Bay & Physics \\
Jason Ho	& Dordt University & Physics \& Engineering \\
Viva Horowitz	& Hamilton College & Physics \\
Hilary Hurst	& San Jose State University & Physics \& Astronomy \\
Martin Kamela	& Elon University & Physics \\
Kishor Kapale	& Western Illinois University & Physics \\
Minjoon Kouh	& Drew University 	& Physics \\
Ronald Kumon	& Kettering University 	& Physics \\
Andres La Rosa	& Portland State University & Physics \\
Paula Lauren	& Lawrence Technological University & Math and Computer Science \\
Thuy Le		& San Jose State University & Electrical Engineering \\
Matthew Leifer	& Chapman University & Physics \\
Kent Leung	& Montclair State University	& Physics \\
Alexandra Liguori& San Francisco State University & Physics \\
Nathan Lundblad & Bates College & Physics and Astronomy \\
Theresa Lynn	& Harvey Mudd College & Physics \\
Viktor Martisovits& Central College & Physics \\
Patrick McQuillan & Chicago Quantum Exchange & Quantum Education\\
Nirav Mehta	& Trinity University & Physics and Astronomy \\
Imran Mirza	& Miami University, Ohio & Physics \\
Brandon Mitchell& West Chester University & Physics \& Engineering \\
Maren Mossman & University of San Diego & Physics and Biophysics \\
Abida Mukarram	& Sacramento State University & Computer Science and Engineering \\
Jorge Munoz	& University of Texas El Paso & Physics \\
Maajida Murdock	& Morgan State University 	& Physics \\
Ricardo Olenewa	& Google	& \\
Daniel Pack	& University of Tennessee at Chattanooga & College of Engineering and Computer Science \\
Gina Passante	& California State University Fullerton & Physics \\
Marlann Patterson& University of Wisconsin -- Stout & Chemistry and Physics \\
Justin Perron 	& California State University San Marcos & Physics \\
Arran Phipps	& California State University East Bay 	& Physics \\
Bissa Quiroz	& California State University Fresno	& Engineering \\
Jamie Raigoza	& California State University Chico	& Computer Science \\
Harindra Rajiyah& California State University Sacramento & Mechanical Engineering \\
Roberto Ramos	& University of the Sciences & Mathematics, Physics, and Statistics \\
Constantin Rasinariu & Loyola University Chicago & Physics \\
Willie Rockward	& Morgan State University & Physics \& Engineering Physics \\
Joshua Sack 	& California State University Long Beach & Mathematics and Statistics \\
J. C. Sanders	& University of Science and Arts of Oklahoma & Physics \\
Heidrun Schmitzer & Xavier University -- Cincinnati, Ohio & Physics \\
Shahed Sharif	& California State University San Marcos & Mathematics \\
Prashant Sharma	& Suffolk University & Physics \\
Paul Stanely	& Beloit College & Physics \\
Ibrahim Sulai	& Bucknell University & Physics and Astronomy \\
Kyle Sundqvist 	& San Diego State University & Physics \\
Maarij Syed	& Rose-Hulman Institute of Technology & Physics \& Optical Engineering \\
Ronald Tackett	& Kettering University & Physics \\
Daniela Topasna	& Virginia Military Institute & Physics and Astronomy \\
%Derrick Tucker	& 
Michael VanValkenburgh	& California State University Sacramento & Mathematics and Statistics \\
%Joel Walsh 
James White	& Juniata College & Physics and Engineering Physics \\
Jeffrey Wolinski& Grove City College & Physics \\
Hiu Yung Wong & San Jose State University & Electrical Engineering \\
Jason Wyenberg & Dordt University & Physics \\
Ahmad Yazdankhah & San Jose State University & Computer Science \\
Guang-Chong Zhu & Lawrence Technological University & Math and Computer Science \\
Todd Zimmerman & University of Wisconsin -- Stout & Chemistry and Physics  \\
\hline
%\end{tabular}
\end{longtable}

\bibliography{QUEST-whitepaper}

\begin{thebibliography}{87}%
\makeatletter
\providecommand \@ifxundefined [1]{%
 \@ifx{#1\undefined}
}%
\providecommand \@ifnum [1]{%
 \ifnum #1\expandafter \@firstoftwo
 \else \expandafter \@secondoftwo
 \fi
}%
\providecommand \@ifx [1]{%
 \ifx #1\expandafter \@firstoftwo
 \else \expandafter \@secondoftwo
 \fi
}%
\providecommand \natexlab [1]{#1}%
\providecommand \enquote  [1]{``#1''}%
\providecommand \bibnamefont  [1]{#1}%
\providecommand \bibfnamefont [1]{#1}%
\providecommand \citenamefont [1]{#1}%
\providecommand \href@noop [0]{\@secondoftwo}%
\providecommand \href [0]{\begingroup \@sanitize@url \@href}%
\providecommand \@href[1]{\@@startlink{#1}\@@href}%
\providecommand \@@href[1]{\endgroup#1\@@endlink}%
\providecommand \@sanitize@url [0]{\catcode `\\12\catcode `\$12\catcode
  `\&12\catcode `\#12\catcode `\^12\catcode `\_12\catcode `\%12\relax}%
\providecommand \@@startlink[1]{}%
\providecommand \@@endlink[0]{}%
\providecommand \url  [0]{\begingroup\@sanitize@url \@url }%
\providecommand \@url [1]{\endgroup\@href {#1}{\urlprefix }}%
\providecommand \urlprefix  [0]{URL }%
\providecommand \Eprint [0]{\href }%
\providecommand \doibase [0]{https://doi.org/}%
\providecommand \selectlanguage [0]{\@gobble}%
\providecommand \bibinfo  [0]{\@secondoftwo}%
\providecommand \bibfield  [0]{\@secondoftwo}%
\providecommand \translation [1]{[#1]}%
\providecommand \BibitemOpen [0]{}%
\providecommand \bibitemStop [0]{}%
\providecommand \bibitemNoStop [0]{.\EOS\space}%
\providecommand \EOS [0]{\spacefactor3000\relax}%
\providecommand \BibitemShut  [1]{\csname bibitem#1\endcsname}%
\let\auto@bib@innerbib\@empty
%</preamble>
\bibitem [{201(2016)}]{2016NSTC}%
  \BibitemOpen
  \href@noop {} {\emph {\bibinfo {title} {Advancing Quantum Information
  Science: National Challenges and Opportunities}}},\ \bibinfo {organization}
  {Committee on Science and Committee on Homeland and National Security}\
  (\bibinfo  {publisher} {National Science and Technology Council, Interagency
  Working Group on Quantum Information Science of the Subcommittee on Physical
  Sciences},\ \bibinfo {year} {2016})\BibitemShut {NoStop}%
\bibitem [{\citenamefont {115th Congress}(2018{\natexlab{a}})}]{NQI}%
  \BibitemOpen
  \bibfield  {author} {\bibinfo {author} {\bibnamefont {115th Congress}},\
  }\href@noop {} {\bibinfo {title} {{H}.{R}. 6227 - {N}ational {Q}uantum
  {I}nitiative {A}ct}} (\bibinfo {year} {2018}{\natexlab{a}}),\ \bibinfo {note}
  {https://www.congress.gov/bill/115th-congress/house-bill/6227/text}\BibitemShut
  {NoStop}%
\bibitem [{\citenamefont {115th
  Congress}(2018{\natexlab{b}})}]{NQI-housereport}%
  \BibitemOpen
  \bibfield  {author} {\bibinfo {author} {\bibnamefont {115th Congress}},\
  }\href@noop {} {\bibinfo {title} {{H}. {R}ept. 115-950 - {N}ational {Q}uantum
  {I}nitiative {A}ct}} (\bibinfo {year} {2018}{\natexlab{b}}),\ \bibinfo {note}
  {https://www.congress.gov/congressional-report/115th-congress/house-report/950}\BibitemShut
  {NoStop}%
\bibitem [{\citenamefont {Cervantes}\ \emph {et~al.}(2021)\citenamefont
  {Cervantes}, \citenamefont {Passante}, \citenamefont {Wilcox},\ and\
  \citenamefont {Pollock}}]{Passante21}%
  \BibitemOpen
  \bibfield  {author} {\bibinfo {author} {\bibfnamefont {B.}~\bibnamefont
  {Cervantes}}, \bibinfo {author} {\bibfnamefont {G.}~\bibnamefont {Passante}},
  \bibinfo {author} {\bibfnamefont {B.}~\bibnamefont {Wilcox}},\ and\ \bibinfo
  {author} {\bibfnamefont {S.}~\bibnamefont {Pollock}},\ }\bibfield  {title}
  {\bibinfo {title} {An overview of quantum information science courses at us
  institutions},\ }\href@noop {} {\bibfield  {journal} {\bibinfo  {journal}
  {2021 Physics Education Research Conference Proceedings}\ }\textbf {\bibinfo
  {volume} {accepted for publication}} (\bibinfo {year} {2021})}\BibitemShut
  {NoStop}%
\bibitem [{QuS(2020)}]{QuSTEAM}%
  \BibitemOpen
  \href@noop {} {\bibinfo {title} {Convergent undergraduate education in
  quantum science, technology, engineering, arts and mathematics}},\ \bibinfo
  {howpublished} {\url{https://qusteam.org/}} (\bibinfo {year}
  {2020})\BibitemShut {NoStop}%
\bibitem [{\citenamefont {Asfaw}\ \emph {et~al.}(2021)\citenamefont {Asfaw},
  \citenamefont {Blais}, \citenamefont {Brown}, \citenamefont {Candelaria},
  \citenamefont {Cantwell}, \citenamefont {Carr}, \citenamefont {Combes},
  \citenamefont {Debroy}, \citenamefont {Donohue}, \citenamefont {Economou},
  \citenamefont {Edwards}, \citenamefont {Fox}, \citenamefont {Girvin},
  \citenamefont {Ho}, \citenamefont {Hurst}, \citenamefont {Jacob},
  \citenamefont {Johnson}, \citenamefont {Johnston-Halperin}, \citenamefont
  {Joynt}, \citenamefont {Kapit}, \citenamefont {Klein-Seetharaman},
  \citenamefont {Laforest}, \citenamefont {Lewandowski}, \citenamefont {Lynn},
  \citenamefont {McRae}, \citenamefont {Merzbacher}, \citenamefont
  {Michalakis}, \citenamefont {Narang}, \citenamefont {Oliver}, \citenamefont
  {Palsberg}, \citenamefont {Pappas}, \citenamefont {Raymer}, \citenamefont
  {Reilly}, \citenamefont {Saffman}, \citenamefont {Searles}, \citenamefont
  {Shapiro},\ and\ \citenamefont {Singh}}]{OSAroadmap}%
  \BibitemOpen
  \bibfield  {author} {\bibinfo {author} {\bibfnamefont {A.}~\bibnamefont
  {Asfaw}}, \bibinfo {author} {\bibfnamefont {A.}~\bibnamefont {Blais}},
  \bibinfo {author} {\bibfnamefont {K.~R.}\ \bibnamefont {Brown}}, \bibinfo
  {author} {\bibfnamefont {J.}~\bibnamefont {Candelaria}}, \bibinfo {author}
  {\bibfnamefont {C.}~\bibnamefont {Cantwell}}, \bibinfo {author}
  {\bibfnamefont {L.~D.}\ \bibnamefont {Carr}}, \bibinfo {author}
  {\bibfnamefont {J.}~\bibnamefont {Combes}}, \bibinfo {author} {\bibfnamefont
  {D.~M.}\ \bibnamefont {Debroy}}, \bibinfo {author} {\bibfnamefont {J.~M.}\
  \bibnamefont {Donohue}}, \bibinfo {author} {\bibfnamefont {S.~E.}\
  \bibnamefont {Economou}}, \bibinfo {author} {\bibfnamefont {E.}~\bibnamefont
  {Edwards}}, \bibinfo {author} {\bibfnamefont {M.~F.~J.}\ \bibnamefont {Fox}},
  \bibinfo {author} {\bibfnamefont {S.~M.}\ \bibnamefont {Girvin}}, \bibinfo
  {author} {\bibfnamefont {A.}~\bibnamefont {Ho}}, \bibinfo {author}
  {\bibfnamefont {H.~M.}\ \bibnamefont {Hurst}}, \bibinfo {author}
  {\bibfnamefont {Z.}~\bibnamefont {Jacob}}, \bibinfo {author} {\bibfnamefont
  {B.~R.}\ \bibnamefont {Johnson}}, \bibinfo {author} {\bibfnamefont
  {E.}~\bibnamefont {Johnston-Halperin}}, \bibinfo {author} {\bibfnamefont
  {R.}~\bibnamefont {Joynt}}, \bibinfo {author} {\bibfnamefont
  {E.}~\bibnamefont {Kapit}}, \bibinfo {author} {\bibfnamefont
  {J.}~\bibnamefont {Klein-Seetharaman}}, \bibinfo {author} {\bibfnamefont
  {M.}~\bibnamefont {Laforest}}, \bibinfo {author} {\bibfnamefont {H.~J.}\
  \bibnamefont {Lewandowski}}, \bibinfo {author} {\bibfnamefont {T.~W.}\
  \bibnamefont {Lynn}}, \bibinfo {author} {\bibfnamefont {C.~R.~H.}\
  \bibnamefont {McRae}}, \bibinfo {author} {\bibfnamefont {C.}~\bibnamefont
  {Merzbacher}}, \bibinfo {author} {\bibfnamefont {S.}~\bibnamefont
  {Michalakis}}, \bibinfo {author} {\bibfnamefont {P.}~\bibnamefont {Narang}},
  \bibinfo {author} {\bibfnamefont {W.~D.}\ \bibnamefont {Oliver}}, \bibinfo
  {author} {\bibfnamefont {J.}~\bibnamefont {Palsberg}}, \bibinfo {author}
  {\bibfnamefont {D.~P.}\ \bibnamefont {Pappas}}, \bibinfo {author}
  {\bibfnamefont {M.~G.}\ \bibnamefont {Raymer}}, \bibinfo {author}
  {\bibfnamefont {D.~J.}\ \bibnamefont {Reilly}}, \bibinfo {author}
  {\bibfnamefont {M.}~\bibnamefont {Saffman}}, \bibinfo {author} {\bibfnamefont
  {T.~A.}\ \bibnamefont {Searles}}, \bibinfo {author} {\bibfnamefont {J.~H.}\
  \bibnamefont {Shapiro}},\ and\ \bibinfo {author} {\bibfnamefont
  {C.}~\bibnamefont {Singh}},\ }\bibfield  {title} {\bibinfo {title} {Building
  a quantum engineering undergraduate program},\ }\href
  {https://arxiv.org/abs/2108.01311v1} {\bibfield  {journal} {\bibinfo
  {journal} {arXiv}\ }\textbf {\bibinfo {volume} {2108.01311}} (\bibinfo {year}
  {2021})}\BibitemShut {NoStop}%
\bibitem [{\citenamefont {Fox}\ \emph {et~al.}(2020)\citenamefont {Fox},
  \citenamefont {Zwickl},\ and\ \citenamefont
  {Lewandowski}}]{fox2020preparing}%
  \BibitemOpen
  \bibfield  {author} {\bibinfo {author} {\bibfnamefont {M.~F.}\ \bibnamefont
  {Fox}}, \bibinfo {author} {\bibfnamefont {B.~M.}\ \bibnamefont {Zwickl}},\
  and\ \bibinfo {author} {\bibfnamefont {H.}~\bibnamefont {Lewandowski}},\
  }\bibfield  {title} {\bibinfo {title} {Preparing for the quantum revolution:
  What is the role of higher education?},\ }\href@noop {} {\bibfield  {journal}
  {\bibinfo  {journal} {Physical Review Physics Education Research}\ }\textbf
  {\bibinfo {volume} {16}},\ \bibinfo {pages} {020131} (\bibinfo {year}
  {2020})}\BibitemShut {NoStop}%
\bibitem [{QED()}]{QEDCjobs}%
  \BibitemOpen
  \href@noop {} {\bibinfo {title} {Quantum economic development consortium
  quantum jobs postings}},\ \bibinfo {howpublished}
  {\url{https://quantumconsortium.org/quantum-jobs/}}\BibitemShut {NoStop}%
\bibitem [{\citenamefont {Loepp}\ and\ \citenamefont {Wootters}(2006)}]{Loepp}%
  \BibitemOpen
  \bibfield  {author} {\bibinfo {author} {\bibfnamefont {S.}~\bibnamefont
  {Loepp}}\ and\ \bibinfo {author} {\bibfnamefont {W.~K.}\ \bibnamefont
  {Wootters}},\ }\href {https://doi.org/10.1017/CBO9780511813719} {\emph
  {\bibinfo {title} {Protecting Information: From Classical Error Correction to
  Quantum Cryptography}}}\ (\bibinfo  {publisher} {Cambridge University
  Press},\ \bibinfo {year} {2006})\BibitemShut {NoStop}%
\bibitem [{\citenamefont {Marrongelle}(2020)}]{DCL21033}%
  \BibitemOpen
  \bibfield  {author} {\bibinfo {author} {\bibfnamefont {K.}~\bibnamefont
  {Marrongelle}},\ }\href@noop {} {\bibinfo {title} {Dear colleague letter:
  Advancing quantum education and workforce development}},\ \bibinfo
  {howpublished} {\url{https://www.nsf.gov/pubs/2021/nsf21033/nsf21033.jsp}}
  (\bibinfo {year} {2020})\BibitemShut {NoStop}%
\bibitem [{\citenamefont {{U.S. Army Contracting Command}}(2021)}]{LQC}%
  \BibitemOpen
  \bibfield  {author} {\bibinfo {author} {\bibnamefont {{U.S. Army Contracting
  Command}}},\ }\href@noop {} {\bibinfo {title} {{LPS Qubit Collaboratory
  (LQC)}}},\ \bibinfo {howpublished}
  {\url{https://www.qubitcollaboratory.org/engage/lqc-open-baa/}} (\bibinfo
  {year} {2021})\BibitemShut {NoStop}%
\bibitem [{Uni()}]{UnitaryFund}%
  \BibitemOpen
  \href@noop {} {\bibinfo {title} {Unitary fund}},\ \bibinfo {howpublished}
  {\url{https://unitary.fund/}}\BibitemShut {NoStop}%
\bibitem [{\citenamefont {{Carol Lynn Alpert et al.}}(2020)}]{KeyConcepts}%
  \BibitemOpen
  \bibfield  {author} {\bibinfo {author} {\bibnamefont {{Carol Lynn Alpert et
  al.}}},\ }\href@noop {} {\bibinfo {title} {Key concepts for future quantum
  information science learners}},\ \bibinfo {howpublished}
  {\url{https://qis-learners.research.illinois.edu/}} (\bibinfo {year}
  {2020}),\ \bibinfo {note} {output from a workshop hosted by the NSF held on
  behalf of the {I}nteragency Working Group on Workforce, Industry and
  Infrastructure under the NSTC Subcommittee on Quantum Information
  Science.}\BibitemShut {Stop}%
\bibitem [{Note1()}]{Note1}%
  \BibitemOpen
  \bibinfo {note} {For a thorough discussion on developing a new course focused
  entirely on QIST readers can refer to ``\protect \emph {Building a Quantum
  Engineering Undergraduate Program}''~\cite {OSAroadmap}}\BibitemShut
  {NoStop}%
\bibitem [{\citenamefont {McIntyre}\ \emph {et~al.}(2012)\citenamefont
  {McIntyre}, \citenamefont {Manogue},\ and\ \citenamefont {Tate}}]{McIntyre}%
  \BibitemOpen
  \bibfield  {author} {\bibinfo {author} {\bibfnamefont {D.~H.}\ \bibnamefont
  {McIntyre}}, \bibinfo {author} {\bibfnamefont {C.~A.}\ \bibnamefont
  {Manogue}},\ and\ \bibinfo {author} {\bibfnamefont {J.}~\bibnamefont
  {Tate}},\ }\href@noop {} {\emph {\bibinfo {title} {Quantum Mechanics: A
  Paradigms Approach}}}\ (\bibinfo  {publisher} {Pearson},\ \bibinfo {year}
  {2012})\BibitemShut {NoStop}%
\bibitem [{\citenamefont {Townsend}(2012)}]{Townsend}%
  \BibitemOpen
  \bibfield  {author} {\bibinfo {author} {\bibfnamefont {J.~S.}\ \bibnamefont
  {Townsend}},\ }\href@noop {} {\emph {\bibinfo {title} {A Modern Approach to
  Quantum Mechanics}}}\ (\bibinfo  {publisher} {University Science Books},\
  \bibinfo {year} {2012})\BibitemShut {NoStop}%
\bibitem [{\citenamefont {Sakurai}\ and\ \citenamefont
  {Napolitano}(2020)}]{Sakurai}%
  \BibitemOpen
  \bibfield  {author} {\bibinfo {author} {\bibfnamefont {J.~J.}\ \bibnamefont
  {Sakurai}}\ and\ \bibinfo {author} {\bibfnamefont {J.}~\bibnamefont
  {Napolitano}},\ }\href {https://doi.org/10.1017/9781108587280} {\emph
  {\bibinfo {title} {Modern Quantum Mechanics}}},\ \bibinfo {edition} {3rd}\
  ed.\ (\bibinfo  {publisher} {Cambridge University Press},\ \bibinfo {year}
  {2020})\BibitemShut {NoStop}%
\bibitem [{\citenamefont {Beck}(2012)}]{Beck}%
  \BibitemOpen
  \bibfield  {author} {\bibinfo {author} {\bibfnamefont {M.}~\bibnamefont
  {Beck}},\ }\href@noop {} {\emph {\bibinfo {title} {Quantum Mechanics: Theory
  and Experiment}}}\ (\bibinfo  {publisher} {Oxford University Press},\
  \bibinfo {year} {2012})\BibitemShut {NoStop}%
\bibitem [{\citenamefont {Zollman}(2016)}]{Zollman16}%
  \BibitemOpen
  \bibfield  {author} {\bibinfo {author} {\bibfnamefont {D.}~\bibnamefont
  {Zollman}},\ }\bibfield  {title} {\bibinfo {title} {Oersted lecture 2014:
  Physics education research and teaching modern modern physics},\ }\href
  {https://doi.org/10.1119/1.4953824} {\bibfield  {journal} {\bibinfo
  {journal} {American Journal of Physics}\ }\textbf {\bibinfo {volume} {84}},\
  \bibinfo {pages} {573} (\bibinfo {year} {2016})},\ \Eprint
  {https://arxiv.org/abs/https://doi.org/10.1119/1.4953824}
  {https://doi.org/10.1119/1.4953824} \BibitemShut {NoStop}%
\bibitem [{\citenamefont {Thorn}\ \emph {et~al.}(2004)\citenamefont {Thorn},
  \citenamefont {Neel}, \citenamefont {Donato}, \citenamefont {Bergreen},
  \citenamefont {Davies},\ and\ \citenamefont {Beck}}]{Thorn04}%
  \BibitemOpen
  \bibfield  {author} {\bibinfo {author} {\bibfnamefont {J.~J.}\ \bibnamefont
  {Thorn}}, \bibinfo {author} {\bibfnamefont {M.~S.}\ \bibnamefont {Neel}},
  \bibinfo {author} {\bibfnamefont {V.~W.}\ \bibnamefont {Donato}}, \bibinfo
  {author} {\bibfnamefont {G.~S.}\ \bibnamefont {Bergreen}}, \bibinfo {author}
  {\bibfnamefont {R.~E.}\ \bibnamefont {Davies}},\ and\ \bibinfo {author}
  {\bibfnamefont {M.}~\bibnamefont {Beck}},\ }\bibfield  {title} {\bibinfo
  {title} {Observing the quantum behavior of light in an undergraduate
  laboratory},\ }\href {https://doi.org/10.1119/1.1737397} {\bibfield
  {journal} {\bibinfo  {journal} {American Journal of Physics}\ }\textbf
  {\bibinfo {volume} {72}},\ \bibinfo {pages} {1210} (\bibinfo {year}
  {2004})},\ \Eprint {https://arxiv.org/abs/https://doi.org/10.1119/1.1737397}
  {https://doi.org/10.1119/1.1737397} \BibitemShut {NoStop}%
\bibitem [{\citenamefont {Carlson}\ \emph {et~al.}(2006)\citenamefont
  {Carlson}, \citenamefont {Olmstead},\ and\ \citenamefont {Beck}}]{Carlson06}%
  \BibitemOpen
  \bibfield  {author} {\bibinfo {author} {\bibfnamefont {J.~A.}\ \bibnamefont
  {Carlson}}, \bibinfo {author} {\bibfnamefont {M.~D.}\ \bibnamefont
  {Olmstead}},\ and\ \bibinfo {author} {\bibfnamefont {M.}~\bibnamefont
  {Beck}},\ }\bibfield  {title} {\bibinfo {title} {Quantum mysteries tested: An
  experiment implementing hardy’s test of local realism},\ }\href
  {https://doi.org/10.1119/1.2167764} {\bibfield  {journal} {\bibinfo
  {journal} {American Journal of Physics}\ }\textbf {\bibinfo {volume} {74}},\
  \bibinfo {pages} {180} (\bibinfo {year} {2006})},\ \Eprint
  {https://arxiv.org/abs/https://doi.org/10.1119/1.2167764}
  {https://doi.org/10.1119/1.2167764} \BibitemShut {NoStop}%
\bibitem [{\citenamefont {Branning}\ \emph {et~al.}(2009)\citenamefont
  {Branning}, \citenamefont {Bhandari},\ and\ \citenamefont
  {Beck}}]{Branning09}%
  \BibitemOpen
  \bibfield  {author} {\bibinfo {author} {\bibfnamefont {D.}~\bibnamefont
  {Branning}}, \bibinfo {author} {\bibfnamefont {S.}~\bibnamefont {Bhandari}},\
  and\ \bibinfo {author} {\bibfnamefont {M.}~\bibnamefont {Beck}},\ }\bibfield
  {title} {\bibinfo {title} {Low-cost coincidence-counting electronics for
  undergraduate quantum optics},\ }\href {https://doi.org/10.1119/1.3116803}
  {\bibfield  {journal} {\bibinfo  {journal} {American Journal of Physics}\
  }\textbf {\bibinfo {volume} {77}},\ \bibinfo {pages} {667} (\bibinfo {year}
  {2009})},\ \Eprint {https://arxiv.org/abs/https://doi.org/10.1119/1.3116803}
  {https://doi.org/10.1119/1.3116803} \BibitemShut {NoStop}%
\bibitem [{\citenamefont {Dederick}\ and\ \citenamefont
  {Beck}(2014)}]{Dederick14}%
  \BibitemOpen
  \bibfield  {author} {\bibinfo {author} {\bibfnamefont {E.}~\bibnamefont
  {Dederick}}\ and\ \bibinfo {author} {\bibfnamefont {M.}~\bibnamefont
  {Beck}},\ }\bibfield  {title} {\bibinfo {title} {Exploring entanglement with
  the help of quantum state measurement},\ }\href
  {https://doi.org/10.1119/1.4883230} {\bibfield  {journal} {\bibinfo
  {journal} {American Journal of Physics}\ }\textbf {\bibinfo {volume} {82}},\
  \bibinfo {pages} {962} (\bibinfo {year} {2014})},\ \Eprint
  {https://arxiv.org/abs/https://doi.org/10.1119/1.4883230}
  {https://doi.org/10.1119/1.4883230} \BibitemShut {NoStop}%
\bibitem [{\citenamefont {Beck}\ and\ \citenamefont {Beck}(2016)}]{Beck16}%
  \BibitemOpen
  \bibfield  {author} {\bibinfo {author} {\bibfnamefont {M.~N.}\ \bibnamefont
  {Beck}}\ and\ \bibinfo {author} {\bibfnamefont {M.}~\bibnamefont {Beck}},\
  }\bibfield  {title} {\bibinfo {title} {Witnessing entanglement in an
  undergraduate laboratory},\ }\href {https://doi.org/10.1119/1.4936623}
  {\bibfield  {journal} {\bibinfo  {journal} {American Journal of Physics}\
  }\textbf {\bibinfo {volume} {84}},\ \bibinfo {pages} {87} (\bibinfo {year}
  {2016})},\ \Eprint {https://arxiv.org/abs/https://doi.org/10.1119/1.4936623}
  {https://doi.org/10.1119/1.4936623} \BibitemShut {NoStop}%
\bibitem [{\citenamefont {Ashby}\ \emph {et~al.}(2016)\citenamefont {Ashby},
  \citenamefont {Schwarz},\ and\ \citenamefont {Schlosshauer}}]{Ashby16}%
  \BibitemOpen
  \bibfield  {author} {\bibinfo {author} {\bibfnamefont {J.~M.}\ \bibnamefont
  {Ashby}}, \bibinfo {author} {\bibfnamefont {P.~D.}\ \bibnamefont {Schwarz}},\
  and\ \bibinfo {author} {\bibfnamefont {M.}~\bibnamefont {Schlosshauer}},\
  }\bibfield  {title} {\bibinfo {title} {Delayed-choice quantum eraser for the
  undergraduate laboratory},\ }\href {https://doi.org/10.1119/1.4938151}
  {\bibfield  {journal} {\bibinfo  {journal} {American Journal of Physics}\
  }\textbf {\bibinfo {volume} {84}},\ \bibinfo {pages} {95} (\bibinfo {year}
  {2016})},\ \Eprint {https://arxiv.org/abs/https://doi.org/10.1119/1.4938151}
  {https://doi.org/10.1119/1.4938151} \BibitemShut {NoStop}%
\bibitem [{\citenamefont {Galvez}\ \emph {et~al.}(2005)\citenamefont {Galvez},
  \citenamefont {Holbrow}, \citenamefont {Pysher}, \citenamefont {Martin},
  \citenamefont {Courtemanche}, \citenamefont {Heilig},\ and\ \citenamefont
  {Spencer}}]{Galvez05}%
  \BibitemOpen
  \bibfield  {author} {\bibinfo {author} {\bibfnamefont {E.~J.}\ \bibnamefont
  {Galvez}}, \bibinfo {author} {\bibfnamefont {C.~H.}\ \bibnamefont {Holbrow}},
  \bibinfo {author} {\bibfnamefont {M.~J.}\ \bibnamefont {Pysher}}, \bibinfo
  {author} {\bibfnamefont {J.~W.}\ \bibnamefont {Martin}}, \bibinfo {author}
  {\bibfnamefont {N.}~\bibnamefont {Courtemanche}}, \bibinfo {author}
  {\bibfnamefont {L.}~\bibnamefont {Heilig}},\ and\ \bibinfo {author}
  {\bibfnamefont {J.}~\bibnamefont {Spencer}},\ }\bibfield  {title} {\bibinfo
  {title} {Interference with correlated photons: Five quantum mechanics
  experiments for undergraduates},\ }\href {https://doi.org/10.1119/1.1796811}
  {\bibfield  {journal} {\bibinfo  {journal} {American Journal of Physics}\
  }\textbf {\bibinfo {volume} {73}},\ \bibinfo {pages} {127} (\bibinfo {year}
  {2005})},\ \Eprint {https://arxiv.org/abs/https://doi.org/10.1119/1.1796811}
  {https://doi.org/10.1119/1.1796811} \BibitemShut {NoStop}%
\bibitem [{\citenamefont {Galvez}(2014)}]{Galvez14}%
  \BibitemOpen
  \bibfield  {author} {\bibinfo {author} {\bibfnamefont {E.~J.}\ \bibnamefont
  {Galvez}},\ }\bibfield  {title} {\bibinfo {title} {Resource letter spe-1:
  Single-photon experiments in the undergraduate laboratory},\ }\href
  {https://doi.org/10.1119/1.4872135} {\bibfield  {journal} {\bibinfo
  {journal} {American Journal of Physics}\ }\textbf {\bibinfo {volume} {82}},\
  \bibinfo {pages} {1018} (\bibinfo {year} {2014})},\ \Eprint
  {https://arxiv.org/abs/https://doi.org/10.1119/1.4872135}
  {https://doi.org/10.1119/1.4872135} \BibitemShut {NoStop}%
\bibitem [{Tea({\natexlab{a}})}]{TeachSpinNMR}%
  \BibitemOpen
  \href@noop {} {\bibinfo {title} {Earth's field nuclear magnetic resonance}},\
  \bibinfo {howpublished} {\url{https://www.teachspin.com/earths-field-nmr}}
  ({\natexlab{a}})\BibitemShut {NoStop}%
\bibitem [{Tea({\natexlab{b}})}]{TeachSpinPNMR}%
  \BibitemOpen
  \href@noop {} {\bibinfo {title} {Pulsed/cw nmr}},\ \bibinfo {howpublished}
  {\url{https://www.teachspin.com/pulsed-nmr}} ({\natexlab{b}})\BibitemShut
  {NoStop}%
\bibitem [{VQO()}]{VQOL}%
  \BibitemOpen
  \href@noop {} {\bibinfo {title} {Virtual quantum optics laboratory}},\
  \bibinfo {howpublished} {\url{https://www.vqol.org/}}\BibitemShut {NoStop}%
\bibitem [{\citenamefont {Candela}(2015)}]{Candela15}%
  \BibitemOpen
  \bibfield  {author} {\bibinfo {author} {\bibfnamefont {D.}~\bibnamefont
  {Candela}},\ }\bibfield  {title} {\bibinfo {title} {Undergraduate
  computational physics projects on quantum computing},\ }\href
  {https://doi.org/10.1119/1.4922296} {\bibfield  {journal} {\bibinfo
  {journal} {American Journal of Physics}\ }\textbf {\bibinfo {volume} {83}},\
  \bibinfo {pages} {688} (\bibinfo {year} {2015})},\ \Eprint
  {https://arxiv.org/abs/https://doi.org/10.1119/1.4922296}
  {https://doi.org/10.1119/1.4922296} \BibitemShut {NoStop}%
\bibitem [{\citenamefont {Galvez}(2021)}]{Galvez21}%
  \BibitemOpen
  \bibfield  {author} {\bibinfo {author} {\bibfnamefont {E.~J.}\ \bibnamefont
  {Galvez}},\ }\bibfield  {title} {\bibinfo {title} {Remote quantum optics
  labs},\ }\href@noop {} {\bibfield  {journal} {\bibinfo  {journal} {Proc.
  SPIE}\ }\textbf {\bibinfo {volume} {11701}} (\bibinfo {year}
  {2021})}\BibitemShut {NoStop}%
\bibitem [{\citenamefont {Asfaw}\ \emph {et~al.}(2020)\citenamefont {Asfaw},
  \citenamefont {Corcoles}, \citenamefont {Bello}, \citenamefont {Ben-Haim},
  \citenamefont {Bozzo-Rey}, \citenamefont {Bravyi}, \citenamefont {Bronn},
  \citenamefont {Capelluto}, \citenamefont {Vazquez}, \citenamefont {Ceroni},
  \citenamefont {Chen}, \citenamefont {Frisch}, \citenamefont {Gambetta},
  \citenamefont {Garion}, \citenamefont {Gil}, \citenamefont {Gonzalez},
  \citenamefont {Harkins}, \citenamefont {Imamichi}, \citenamefont {Kang},
  \citenamefont {h.~Karamlou}, \citenamefont {Loredo}, \citenamefont {McKay},
  \citenamefont {Mezzacapo}, \citenamefont {Minev}, \citenamefont {Movassagh},
  \citenamefont {Nannicini}, \citenamefont {Nation}, \citenamefont {Phan},
  \citenamefont {Pistoia}, \citenamefont {Rattew}, \citenamefont {Schaefer},
  \citenamefont {Shabani}, \citenamefont {Smolin}, \citenamefont {Stenger},
  \citenamefont {Temme}, \citenamefont {Tod}, \citenamefont {Wood},\ and\
  \citenamefont {Wootton.}}]{Qiskit}%
  \BibitemOpen
  \bibfield  {author} {\bibinfo {author} {\bibfnamefont {A.}~\bibnamefont
  {Asfaw}}, \bibinfo {author} {\bibfnamefont {A.}~\bibnamefont {Corcoles}},
  \bibinfo {author} {\bibfnamefont {L.}~\bibnamefont {Bello}}, \bibinfo
  {author} {\bibfnamefont {Y.}~\bibnamefont {Ben-Haim}}, \bibinfo {author}
  {\bibfnamefont {M.}~\bibnamefont {Bozzo-Rey}}, \bibinfo {author}
  {\bibfnamefont {S.}~\bibnamefont {Bravyi}}, \bibinfo {author} {\bibfnamefont
  {N.}~\bibnamefont {Bronn}}, \bibinfo {author} {\bibfnamefont
  {L.}~\bibnamefont {Capelluto}}, \bibinfo {author} {\bibfnamefont {A.~C.}\
  \bibnamefont {Vazquez}}, \bibinfo {author} {\bibfnamefont {J.}~\bibnamefont
  {Ceroni}}, \bibinfo {author} {\bibfnamefont {R.}~\bibnamefont {Chen}},
  \bibinfo {author} {\bibfnamefont {A.}~\bibnamefont {Frisch}}, \bibinfo
  {author} {\bibfnamefont {J.}~\bibnamefont {Gambetta}}, \bibinfo {author}
  {\bibfnamefont {S.}~\bibnamefont {Garion}}, \bibinfo {author} {\bibfnamefont
  {L.}~\bibnamefont {Gil}}, \bibinfo {author} {\bibfnamefont {S.~D. L.~P.}\
  \bibnamefont {Gonzalez}}, \bibinfo {author} {\bibfnamefont {F.}~\bibnamefont
  {Harkins}}, \bibinfo {author} {\bibfnamefont {T.}~\bibnamefont {Imamichi}},
  \bibinfo {author} {\bibfnamefont {H.}~\bibnamefont {Kang}}, \bibinfo {author}
  {\bibfnamefont {A.}~\bibnamefont {h.~Karamlou}}, \bibinfo {author}
  {\bibfnamefont {R.}~\bibnamefont {Loredo}}, \bibinfo {author} {\bibfnamefont
  {D.}~\bibnamefont {McKay}}, \bibinfo {author} {\bibfnamefont
  {A.}~\bibnamefont {Mezzacapo}}, \bibinfo {author} {\bibfnamefont
  {Z.}~\bibnamefont {Minev}}, \bibinfo {author} {\bibfnamefont
  {R.}~\bibnamefont {Movassagh}}, \bibinfo {author} {\bibfnamefont
  {G.}~\bibnamefont {Nannicini}}, \bibinfo {author} {\bibfnamefont
  {P.}~\bibnamefont {Nation}}, \bibinfo {author} {\bibfnamefont
  {A.}~\bibnamefont {Phan}}, \bibinfo {author} {\bibfnamefont {M.}~\bibnamefont
  {Pistoia}}, \bibinfo {author} {\bibfnamefont {A.}~\bibnamefont {Rattew}},
  \bibinfo {author} {\bibfnamefont {J.}~\bibnamefont {Schaefer}}, \bibinfo
  {author} {\bibfnamefont {J.}~\bibnamefont {Shabani}}, \bibinfo {author}
  {\bibfnamefont {J.}~\bibnamefont {Smolin}}, \bibinfo {author} {\bibfnamefont
  {J.}~\bibnamefont {Stenger}}, \bibinfo {author} {\bibfnamefont
  {K.}~\bibnamefont {Temme}}, \bibinfo {author} {\bibfnamefont
  {M.}~\bibnamefont {Tod}}, \bibinfo {author} {\bibfnamefont {S.}~\bibnamefont
  {Wood}},\ and\ \bibinfo {author} {\bibfnamefont {J.}~\bibnamefont
  {Wootton.}},\ }\href {http://community.qiskit.org/textbook} {\bibinfo {title}
  {Learn quantum computation using qiskit}} (\bibinfo {year}
  {2020})\BibitemShut {NoStop}%
\bibitem [{Cir()}]{Cirq}%
  \BibitemOpen
  \href {https://quantumai.google/cirq/google/concepts} {\bibinfo {title}
  {Google quantum computing service}}\BibitemShut {NoStop}%
\bibitem [{Qsh()}]{Qsharp}%
  \BibitemOpen
  \href
  {https://azure.microsoft.com/en-us/resources/development-kit/quantum-computing/}
  {\bibinfo {title} {Q\# and the quantum development kit}}\BibitemShut
  {NoStop}%
\bibitem [{AWS()}]{AWS}%
  \BibitemOpen
  \href {https://aws.amazon.com/braket/} {\bibinfo {title} {Amazon
  braket}}\BibitemShut {NoStop}%
\bibitem [{QCd()}]{QCdotcom}%
  \BibitemOpen
  \href@noop {} {\bibinfo {title} {Strangeworks {Q}{C}}},\ \bibinfo
  {howpublished} {\url{https://quantumcomputing.com/}}\BibitemShut {NoStop}%
\bibitem [{Note2()}]{Note2}%
  \BibitemOpen
  \bibinfo {note} {The mention of these products does not constitute an
  endorsement by the authors or workshop attendees but rather a brief list of
  products we are aware of. We anticipate the number of products in this space
  will increase steadily in the coming years}\BibitemShut {NoStop}%
\bibitem [{Qub()}]{Qubitekk}%
  \BibitemOpen
  \href@noop {} {\bibinfo {title} {Qubitekk: Quantum mechanics lab kit}},\
  \bibinfo {howpublished}
  {\url{http://qubitekk.com/products/quantum-mechanics-lab-kit/}}\BibitemShut
  {NoStop}%
\bibitem [{Spi()}]{Spinflex}%
  \BibitemOpen
  \href@noop {} {\bibinfo {title} {Spinedu: Educational kit based on
  solid-state spin qubits in diamond for quantum technology}},\ \bibinfo
  {howpublished} {\url{https://spin-flex.com/spinedu/}}\BibitemShut {NoStop}%
\bibitem [{qut()}]{qutools}%
  \BibitemOpen
  \href@noop {} {\bibinfo {title} {qutools: Quantum physics kits}},\ \bibinfo
  {howpublished}
  {\url{https://qutools.com/quantum-physics-education-science-kits/}}\BibitemShut
  {NoStop}%
\bibitem [{Alp()}]{Alpha}%
  \BibitemOpen
  \href@noop {} {\bibinfo {title} {{ALPhA Advanced Laboratories}}},\ \bibinfo
  {howpublished} {\url{https://advlab.org/}}\BibitemShut {NoStop}%
\bibitem [{\citenamefont {Schumacher}\ and\ \citenamefont
  {Westmoreland}(2010)}]{QPSI}%
  \BibitemOpen
  \bibfield  {author} {\bibinfo {author} {\bibfnamefont {B.}~\bibnamefont
  {Schumacher}}\ and\ \bibinfo {author} {\bibfnamefont {M.}~\bibnamefont
  {Westmoreland}},\ }\href {https://doi.org/10.1017/CBO9780511814006} {\emph
  {\bibinfo {title} {Quantum Processes Systems, and Information}}}\ (\bibinfo
  {publisher} {Cambridge University Press},\ \bibinfo {year}
  {2010})\BibitemShut {NoStop}%
\bibitem [{\citenamefont {{National Academies of Sciences, Engineering, and
  Medicine}}(2019{\natexlab{a}})}]{ProgressAndProspects}%
  \BibitemOpen
  \bibfield  {author} {\bibinfo {author} {\bibnamefont {{National Academies of
  Sciences, Engineering, and Medicine}}},\ }\href
  {https://doi.org/10.17226/25196} {\emph {\bibinfo {title} {Quantum Computing:
  Progress and Prospects}}},\ edited by\ \bibinfo {editor} {\bibfnamefont
  {E.}~\bibnamefont {Grumbling}}\ and\ \bibinfo {editor} {\bibfnamefont
  {M.}~\bibnamefont {Horowitz}}\ (\bibinfo  {publisher} {The National Academies
  Press},\ \bibinfo {address} {Washington, DC},\ \bibinfo {year}
  {2019})\BibitemShut {NoStop}%
\bibitem [{\citenamefont {Lipton}\ and\ \citenamefont
  {Regan}(2021)}]{LiptonRegan2021}%
  \BibitemOpen
  \bibfield  {author} {\bibinfo {author} {\bibfnamefont {R.~J.}\ \bibnamefont
  {Lipton}}\ and\ \bibinfo {author} {\bibfnamefont {K.~W.}\ \bibnamefont
  {Regan}},\ }\href@noop {} {\emph {\bibinfo {title} {Quantum Algorithms via
  Linear Algebra: A Primer}}}\ (\bibinfo  {publisher} {MIT Press},\ \bibinfo
  {year} {2021})\BibitemShut {NoStop}%
\bibitem [{\citenamefont {Childs}\ and\ \citenamefont {van
  Dam}(2010)}]{algebra-problems}%
  \BibitemOpen
  \bibfield  {author} {\bibinfo {author} {\bibfnamefont {A.~M.}\ \bibnamefont
  {Childs}}\ and\ \bibinfo {author} {\bibfnamefont {W.}~\bibnamefont {van
  Dam}},\ }\bibfield  {title} {\bibinfo {title} {Quantum algorithms for
  algebraic problems},\ }\href {https://doi.org/10.1103/revmodphys.82.1}
  {\bibfield  {journal} {\bibinfo  {journal} {Reviews of Modern Physics}\
  }\textbf {\bibinfo {volume} {82}},\ \bibinfo {pages} {1–52} (\bibinfo
  {year} {2010})}\BibitemShut {NoStop}%
\bibitem [{\citenamefont {Abramsky}(2007)}]{Abramsky2007}%
  \BibitemOpen
  \bibfield  {author} {\bibinfo {author} {\bibfnamefont {S.}~\bibnamefont
  {Abramsky}},\ }\bibfield  {title} {\bibinfo {title} {Temperley–lieb
  algebra: From knot theory to logic and computation via quantum mechanics},\
  }in\ \href@noop {} {\emph {\bibinfo {booktitle} {Mathematics of quantum
  computation and quantum technology}}},\ \bibinfo {editor} {edited by\
  \bibinfo {editor} {\bibfnamefont {L.}~\bibnamefont {Kauffman}}\ and\ \bibinfo
  {editor} {\bibfnamefont {S.~J.}\ \bibnamefont {Lomonaco}}}\ (\bibinfo {year}
  {2007})\BibitemShut {NoStop}%
\bibitem [{\citenamefont {Coecke}\ and\ \citenamefont
  {Kissinger}(2017)}]{CoeckeKissenger2017}%
  \BibitemOpen
  \bibfield  {author} {\bibinfo {author} {\bibfnamefont {B.}~\bibnamefont
  {Coecke}}\ and\ \bibinfo {author} {\bibfnamefont {A.}~\bibnamefont
  {Kissinger}},\ }\href {https://doi.org/https://doi.org/10.1017/9781316219317}
  {\emph {\bibinfo {title} {Picturing Quantum Processes}}}\ (\bibinfo
  {publisher} {Cambridge University Press},\ \bibinfo {year}
  {2017})\BibitemShut {NoStop}%
\bibitem [{\citenamefont {Kalmbach}(1983)}]{Kalmbach1983}%
  \BibitemOpen
  \bibfield  {author} {\bibinfo {author} {\bibfnamefont {G.}~\bibnamefont
  {Kalmbach}},\ }\href@noop {} {\emph {\bibinfo {title} {Orthomodular
  Lattices}}}\ (\bibinfo  {publisher} {Academic Press},\ \bibinfo {year}
  {1983})\BibitemShut {NoStop}%
\bibitem [{\citenamefont {Harding}(2007)}]{Harding2007}%
  \BibitemOpen
  \bibfield  {author} {\bibinfo {author} {\bibfnamefont {J.}~\bibnamefont
  {Harding}},\ }\bibfield  {title} {\bibinfo {title} {- the source of the
  orthomodular law},\ }in\ \href
  {https://doi.org/https://doi.org/10.1016/B978-044452870-4/50035-2} {\emph
  {\bibinfo {booktitle} {Handbook of Quantum Logic and Quantum Structures}}},\
  \bibinfo {editor} {edited by\ \bibinfo {editor} {\bibfnamefont
  {K.}~\bibnamefont {Engesser}}, \bibinfo {editor} {\bibfnamefont {D.~M.}\
  \bibnamefont {Gabbay}},\ and\ \bibinfo {editor} {\bibfnamefont
  {D.}~\bibnamefont {Lehmann}}}\ (\bibinfo  {publisher} {Elsevier Science
  B.V.},\ \bibinfo {address} {Amsterdam},\ \bibinfo {year} {2007})\ pp.\
  \bibinfo {pages} {555--586}\BibitemShut {NoStop}%
\bibitem [{\citenamefont {Piron}(1976)}]{Piron1976}%
  \BibitemOpen
  \bibfield  {author} {\bibinfo {author} {\bibfnamefont {C.}~\bibnamefont
  {Piron}},\ }\href@noop {} {\emph {\bibinfo {title} {Foundations of Quantum
  Physics}}}\ (\bibinfo  {publisher} {W.A. Benjamin Inc.},\ \bibinfo {year}
  {1976})\BibitemShut {NoStop}%
\bibitem [{\citenamefont {Chiara}\ \emph {et~al.}(2004)\citenamefont {Chiara},
  \citenamefont {Giuntini},\ and\ \citenamefont
  {Greechie}}]{DallaChiaraGiuntiniGreechie2004}%
  \BibitemOpen
  \bibfield  {author} {\bibinfo {author} {\bibfnamefont {M.~L.~D.}\
  \bibnamefont {Chiara}}, \bibinfo {author} {\bibfnamefont {R.}~\bibnamefont
  {Giuntini}},\ and\ \bibinfo {author} {\bibfnamefont {R.}~\bibnamefont
  {Greechie}},\ }\href@noop {} {\emph {\bibinfo {title} {Reasoning in Quantum
  Theory: Sharp and Unsharp Quantum Logics}}}\ (\bibinfo  {publisher} {Springer
  Netherlands},\ \bibinfo {year} {2004})\BibitemShut {NoStop}%
\bibitem [{\citenamefont {Rosenthal}(1990)}]{Rosenthal1990}%
  \BibitemOpen
  \bibfield  {author} {\bibinfo {author} {\bibfnamefont {K.}~\bibnamefont
  {Rosenthal}},\ }\href@noop {} {\emph {\bibinfo {title} {{Quantales and Their
  Applications}}}}\ (\bibinfo  {publisher} {Longman Higher Education},\
  \bibinfo {year} {1990})\BibitemShut {NoStop}%
\bibitem [{\citenamefont {Abramsky}\ and\ \citenamefont
  {Vickers}(1993)}]{AbramskyVickers1993}%
  \BibitemOpen
  \bibfield  {author} {\bibinfo {author} {\bibfnamefont {S.}~\bibnamefont
  {Abramsky}}\ and\ \bibinfo {author} {\bibfnamefont {S.}~\bibnamefont
  {Vickers}},\ }\bibfield  {title} {\bibinfo {title} {Quantales, observational
  logic and process semantics},\ }\href
  {https://doi.org/10.1017/S0960129500000189} {\bibfield  {journal} {\bibinfo
  {journal} {Mathematical Structures in Computer Science}\ }\textbf {\bibinfo
  {volume} {3}},\ \bibinfo {pages} {161} (\bibinfo {year} {1993})}\BibitemShut
  {NoStop}%
\bibitem [{\citenamefont {Herrmann}\ and\ \citenamefont
  {Ziegler}(2016)}]{HerrmannZiegler2016}%
  \BibitemOpen
  \bibfield  {author} {\bibinfo {author} {\bibfnamefont {C.}~\bibnamefont
  {Herrmann}}\ and\ \bibinfo {author} {\bibfnamefont {M.}~\bibnamefont
  {Ziegler}},\ }\bibfield  {title} {\bibinfo {title} {Computational complexity
  of quantum satisfiability},\ }\href {https://doi.org/10.1145/2869073}
  {\bibfield  {journal} {\bibinfo  {journal} {Journal of the ACM}\ }\textbf
  {\bibinfo {volume} {63}},\ \bibinfo {pages} {1} (\bibinfo {year}
  {2016})}\BibitemShut {NoStop}%
\bibitem [{\citenamefont {Baltag}\ and\ \citenamefont
  {Smets}(2006)}]{Baltag2006}%
  \BibitemOpen
  \bibfield  {author} {\bibinfo {author} {\bibfnamefont {A.}~\bibnamefont
  {Baltag}}\ and\ \bibinfo {author} {\bibfnamefont {S.}~\bibnamefont {Smets}},\
  }\bibfield  {title} {\bibinfo {title} {{LQP: The Dynamic Logic of Quantum
  Information}},\ }\href@noop {} {\bibfield  {journal} {\bibinfo  {journal}
  {Mathematical Structures in Computer Science}\ }\textbf {\bibinfo {volume}
  {16}},\ \bibinfo {pages} {491} (\bibinfo {year} {2006})}\BibitemShut
  {NoStop}%
\bibitem [{\citenamefont {Bergfeld}\ and\ \citenamefont
  {Sack}(2017)}]{Bergfeld2017}%
  \BibitemOpen
  \bibfield  {author} {\bibinfo {author} {\bibfnamefont {J.}~\bibnamefont
  {Bergfeld}}\ and\ \bibinfo {author} {\bibfnamefont {J.}~\bibnamefont
  {Sack}},\ }\bibfield  {title} {\bibinfo {title} {Deriving the correctness of
  quantum protocols in the probabilistic logic for quantum programs},\ }\href
  {https://doi.org/10.1007/s00500-015-1802-6} {\bibfield  {journal} {\bibinfo
  {journal} {Soft Computing}\ }\textbf {\bibinfo {volume} {21}},\ \bibinfo
  {pages} {pages 1421–1441} (\bibinfo {year} {2017})}\BibitemShut {NoStop}%
\bibitem [{\citenamefont {Abramsky}\ and\ \citenamefont
  {Hardy}(2012)}]{AbramskyHardy2012}%
  \BibitemOpen
  \bibfield  {author} {\bibinfo {author} {\bibfnamefont {S.}~\bibnamefont
  {Abramsky}}\ and\ \bibinfo {author} {\bibfnamefont {L.}~\bibnamefont
  {Hardy}},\ }\bibfield  {title} {\bibinfo {title} {Logical bell
  inequalities},\ }\href {https://doi.org/10.1103/PhysRevA.85.062114}
  {\bibfield  {journal} {\bibinfo  {journal} {Phys. Rev. A}\ }\textbf {\bibinfo
  {volume} {85}},\ \bibinfo {pages} {062114} (\bibinfo {year}
  {2012})}\BibitemShut {NoStop}%
\bibitem [{\citenamefont {Wilce}(2017)}]{sep-qt-quantlog}%
  \BibitemOpen
  \bibfield  {author} {\bibinfo {author} {\bibfnamefont {A.}~\bibnamefont
  {Wilce}},\ }\bibfield  {title} {\bibinfo {title} {{Quantum Logic and
  Probability Theory}},\ }in\ \href@noop {} {\emph {\bibinfo {booktitle} {The
  {Stanford} Encyclopedia of Philosophy}}},\ \bibinfo {editor} {edited by\
  \bibinfo {editor} {\bibfnamefont {E.~N.}\ \bibnamefont {Zalta}}}\ (\bibinfo
  {publisher} {Metaphysics Research Lab, Stanford University},\ \bibinfo {year}
  {2017})\ \bibinfo {edition} {{S}pring 2017}\ ed.\BibitemShut {Stop}%
\bibitem [{\citenamefont {Bergfeld}\ \emph {et~al.}(2015)\citenamefont
  {Bergfeld}, \citenamefont {Kishida}, \citenamefont {Sack},\ and\
  \citenamefont {Zhong}}]{BergfeldEtAl2015}%
  \BibitemOpen
  \bibfield  {author} {\bibinfo {author} {\bibfnamefont {J.~M.}\ \bibnamefont
  {Bergfeld}}, \bibinfo {author} {\bibfnamefont {K.}~\bibnamefont {Kishida}},
  \bibinfo {author} {\bibfnamefont {J.}~\bibnamefont {Sack}},\ and\ \bibinfo
  {author} {\bibfnamefont {S.}~\bibnamefont {Zhong}},\ }\bibfield  {title}
  {\bibinfo {title} {Duality for the logic of quantum actions},\ }\href
  {https://doi.org/10.1007/s11225-014-9592-x} {\bibfield  {journal} {\bibinfo
  {journal} {Studia Logica}\ }\textbf {\bibinfo {volume} {103}},\ \bibinfo
  {pages} {781–805} (\bibinfo {year} {2015})}\BibitemShut {NoStop}%
\bibitem [{\citenamefont {Coecke}\ and\ \citenamefont
  {Smets}(2004)}]{CoeckeSmets2004}%
  \BibitemOpen
  \bibfield  {author} {\bibinfo {author} {\bibfnamefont {B.}~\bibnamefont
  {Coecke}}\ and\ \bibinfo {author} {\bibfnamefont {S.}~\bibnamefont {Smets}},\
  }\bibfield  {title} {\bibinfo {title} {The sasaki hook is not a [static]
  implicative connective but induces a backward [in time] dynamic one that
  assigns causes},\ }\href {https://doi.org/10.1023/B:IJTP.0000048815.92983.6e}
  {\bibfield  {journal} {\bibinfo  {journal} {International Journal of
  Theoretical Physics}\ }\textbf {\bibinfo {volume} {43}},\ \bibinfo {pages}
  {1705} (\bibinfo {year} {2004})}\BibitemShut {NoStop}%
\bibitem [{\citenamefont {Nielsen}\ and\ \citenamefont
  {Chuang}(2011)}]{Nielsen2011}%
  \BibitemOpen
  \bibfield  {author} {\bibinfo {author} {\bibfnamefont {M.~A.}\ \bibnamefont
  {Nielsen}}\ and\ \bibinfo {author} {\bibfnamefont {I.~L.}\ \bibnamefont
  {Chuang}},\ }\href@noop {} {\emph {\bibinfo {title} {Quantum Computation and
  Quantum Information}}}\ (\bibinfo  {publisher} {Cambridge University Press},\
  \bibinfo {year} {2011})\BibitemShut {NoStop}%
\bibitem [{\citenamefont {SELINGER}(2004)}]{selinger_2004}%
  \BibitemOpen
  \bibfield  {author} {\bibinfo {author} {\bibfnamefont {P.}~\bibnamefont
  {SELINGER}},\ }\bibfield  {title} {\bibinfo {title} {Towards a quantum
  programming language},\ }\href {https://doi.org/10.1017/S0960129504004256}
  {\bibfield  {journal} {\bibinfo  {journal} {Mathematical Structures in
  Computer Science}\ }\textbf {\bibinfo {volume} {14}},\ \bibinfo {pages}
  {527–586} (\bibinfo {year} {2004})}\BibitemShut {NoStop}%
\bibitem [{\citenamefont {Eisenberg}\ \emph {et~al.}(2019)\citenamefont
  {Eisenberg}, \citenamefont {Green}, \citenamefont {Lumsdaine}, \citenamefont
  {Kim}, \citenamefont {Mau}, \citenamefont {Mohan}, \citenamefont {Ng},
  \citenamefont {Ravelomanantsoa-Ratsimihah}, \citenamefont {Ross},
  \citenamefont {Scherer}, \citenamefont {Selinger}, \citenamefont {Valiron},
  \citenamefont {Virodov},\ and\ \citenamefont {Zdancewic}}]{quipper}%
  \BibitemOpen
  \bibfield  {author} {\bibinfo {author} {\bibfnamefont {R.}~\bibnamefont
  {Eisenberg}}, \bibinfo {author} {\bibfnamefont {A.~S.}\ \bibnamefont
  {Green}}, \bibinfo {author} {\bibfnamefont {P.~L.}\ \bibnamefont
  {Lumsdaine}}, \bibinfo {author} {\bibfnamefont {K.}~\bibnamefont {Kim}},
  \bibinfo {author} {\bibfnamefont {S.-C.}\ \bibnamefont {Mau}}, \bibinfo
  {author} {\bibfnamefont {B.}~\bibnamefont {Mohan}}, \bibinfo {author}
  {\bibfnamefont {W.}~\bibnamefont {Ng}}, \bibinfo {author} {\bibfnamefont
  {J.}~\bibnamefont {Ravelomanantsoa-Ratsimihah}}, \bibinfo {author}
  {\bibfnamefont {N.~J.}\ \bibnamefont {Ross}}, \bibinfo {author}
  {\bibfnamefont {A.}~\bibnamefont {Scherer}}, \bibinfo {author} {\bibfnamefont
  {P.}~\bibnamefont {Selinger}}, \bibinfo {author} {\bibfnamefont
  {B.}~\bibnamefont {Valiron}}, \bibinfo {author} {\bibfnamefont
  {A.}~\bibnamefont {Virodov}},\ and\ \bibinfo {author} {\bibfnamefont {S.~A.}\
  \bibnamefont {Zdancewic}},\ }\href@noop {} {\bibinfo {title} {The quipper
  language}},\ \bibinfo {howpublished}
  {\url{https://www.mathstat.dal.ca/~selinger/quipper/}} (\bibinfo {year}
  {2019})\BibitemShut {NoStop}%
\bibitem [{\citenamefont {Green}\ \emph {et~al.}(2013)\citenamefont {Green},
  \citenamefont {Lumsdaine}, \citenamefont {Ross}, \citenamefont {Selinger},\
  and\ \citenamefont {Valiron}}]{quippertut}%
  \BibitemOpen
  \bibfield  {author} {\bibinfo {author} {\bibfnamefont {A.~S.}\ \bibnamefont
  {Green}}, \bibinfo {author} {\bibfnamefont {P.~L.}\ \bibnamefont
  {Lumsdaine}}, \bibinfo {author} {\bibfnamefont {N.~J.}\ \bibnamefont {Ross}},
  \bibinfo {author} {\bibfnamefont {P.}~\bibnamefont {Selinger}},\ and\
  \bibinfo {author} {\bibfnamefont {B.}~\bibnamefont {Valiron}},\ }\bibfield
  {title} {\bibinfo {title} {An introduction to quantum programming in
  quipper},\ }\href {https://doi.org/10.1007/978-3-642-38986-3_10} {\bibfield
  {journal} {\bibinfo  {journal} {Lecture Notes in Computer Science}\ ,\
  \bibinfo {pages} {110 }} (\bibinfo {year} {2013})}\BibitemShut {NoStop}%
\bibitem [{\citenamefont {Bichsel}\ \emph {et~al.}()\citenamefont {Bichsel},
  \citenamefont {Baader}, \citenamefont {Gehr},\ and\ \citenamefont
  {Vechev}}]{silq}%
  \BibitemOpen
  \bibfield  {author} {\bibinfo {author} {\bibfnamefont {B.}~\bibnamefont
  {Bichsel}}, \bibinfo {author} {\bibfnamefont {M.}~\bibnamefont {Baader}},
  \bibinfo {author} {\bibfnamefont {T.}~\bibnamefont {Gehr}},\ and\ \bibinfo
  {author} {\bibfnamefont {M.}~\bibnamefont {Vechev}},\ }\href@noop {}
  {\bibinfo {title} {Silq}},\ \bibinfo {howpublished}
  {\url{https://silq.ethz.ch/}}\BibitemShut {NoStop}%
\bibitem [{Com()}]{ComplexityZoo}%
  \BibitemOpen
  \href@noop {} {\bibinfo {title} {Complexity zoo}},\ \bibinfo {howpublished}
  {\url{https://complexityzoo.net/Complexity_Zoo}}\BibitemShut {NoStop}%
\bibitem [{DoD()}]{DoDSTEM}%
  \BibitemOpen
  \href@noop {} {\bibinfo {title} {{D}o{D} stem opportunities}},\ \bibinfo
  {howpublished}
  {\url{https://www.dodstem.us/participate/opportunities/}}\BibitemShut
  {NoStop}%
\bibitem [{ORN()}]{ORNLmail}%
  \BibitemOpen
  \href@noop {} {\bibinfo {title} {The quantum computing institute external
  mailman group}},\ \bibinfo {howpublished}
  {\url{https://elist.ornl.gov/mailman/listinfo/qci-external}}\BibitemShut
  {NoStop}%
\bibitem [{Phy()}]{PhysAdopt}%
  \BibitemOpen
  \href@noop {} {\bibinfo {title} {Adopt-a-physicist}},\ \bibinfo
  {howpublished} {\url{https://www.adoptaphysicist.org/}}\BibitemShut {NoStop}%
\bibitem [{\citenamefont {{The AIP national task force to elevate african
  american representation in undergraduate physics \& astronomy
  (TEAM-UP)}}(2020)}]{Teamup}%
  \BibitemOpen
  \bibfield  {author} {\bibinfo {author} {\bibnamefont {{The AIP national task
  force to elevate african american representation in undergraduate physics \&
  astronomy (TEAM-UP)}}},\ }\href@noop {} {\emph {\bibinfo {title} {The Time is
  Now: Systemic Changes to Increase African Americans with Bachelor's Degrees
  in Physics and Astronomy}}}\ (\bibinfo  {publisher} {American Institute of
  Physics},\ \bibinfo {address} {College Park, MD},\ \bibinfo {year}
  {2020})\BibitemShut {NoStop}%
\bibitem [{\citenamefont {Hilborn}\ \emph {et~al.}(2003)\citenamefont
  {Hilborn}, \citenamefont {Howes},\ and\ \citenamefont {Krane}}]{Spinup}%
  \BibitemOpen
  \bibinfo {editor} {\bibfnamefont {R.~C.}\ \bibnamefont {Hilborn}}, \bibinfo
  {editor} {\bibfnamefont {R.~H.}\ \bibnamefont {Howes}},\ and\ \bibinfo
  {editor} {\bibfnamefont {K.~S.}\ \bibnamefont {Krane}},\ eds.,\ \href@noop {}
  {\emph {\bibinfo {title} {Strategic Programs for Innovations in Undergraduate
  Physics: Project Report}}}\ (\bibinfo  {publisher} {The American Association
  of Physics Teachers},\ \bibinfo {address} {College Park, MD},\ \bibinfo
  {year} {2003})\BibitemShut {NoStop}%
\bibitem [{\citenamefont {Graham}\ \emph {et~al.}(2013)\citenamefont {Graham},
  \citenamefont {Frederick}, \citenamefont {Byars-Winston}, \citenamefont
  {Hunter},\ and\ \citenamefont {Handelsman}}]{graham2013increasing}%
  \BibitemOpen
  \bibfield  {author} {\bibinfo {author} {\bibfnamefont {M.~J.}\ \bibnamefont
  {Graham}}, \bibinfo {author} {\bibfnamefont {J.}~\bibnamefont {Frederick}},
  \bibinfo {author} {\bibfnamefont {A.}~\bibnamefont {Byars-Winston}}, \bibinfo
  {author} {\bibfnamefont {A.-B.}\ \bibnamefont {Hunter}},\ and\ \bibinfo
  {author} {\bibfnamefont {J.}~\bibnamefont {Handelsman}},\ }\bibfield  {title}
  {\bibinfo {title} {Increasing persistence of college students in stem},\
  }\href@noop {} {\bibfield  {journal} {\bibinfo  {journal} {Science}\ }\textbf
  {\bibinfo {volume} {341}},\ \bibinfo {pages} {1455} (\bibinfo {year}
  {2013})},\ \bibinfo {note} {and references 13-16 therein.}\BibitemShut
  {Stop}%
\bibitem [{\citenamefont {{The American Physical Society (APS) and the American
  Association of Physics Teachers (AAPT)}}()}]{EP3}%
  \BibitemOpen
  \bibfield  {author} {\bibinfo {author} {\bibnamefont {{The American Physical
  Society (APS) and the American Association of Physics Teachers (AAPT)}}},\
  }\href@noop {} {\bibinfo {title} {Effective practices for physics programs
  (ep3)}},\ \bibinfo {howpublished} {\url{https://ep3guide.org/}}\BibitemShut
  {NoStop}%
\bibitem [{VIP()}]{VIP}%
  \BibitemOpen
  \href@noop {} {\bibinfo {title} {Valley industry partnership for cooperative
  education}},\ \bibinfo {howpublished}
  {\url{http://fresnostate.edu/engineering/vip/}}\BibitemShut {NoStop}%
\bibitem [{axi()}]{axioms}%
  \BibitemOpen
  \href@noop {} {}\bibinfo {howpublished}
  {\url{http://math.sfsu.edu/federico/}}\BibitemShut {NoStop}%
\bibitem [{\citenamefont {Feder}(2017)}]{Feder17}%
  \BibitemOpen
  \bibfield  {author} {\bibinfo {author} {\bibfnamefont {T.}~\bibnamefont
  {Feder}},\ }\bibfield  {title} {\bibinfo {title} {College-level project-based
  learning gains popularity},\ }\href {https://doi.org/10.1063/PT.3.3589}
  {\bibfield  {journal} {\bibinfo  {journal} {Physics Today}\ }\textbf
  {\bibinfo {volume} {70}},\ \bibinfo {pages} {28} (\bibinfo {year} {2017})},\
  \Eprint {https://arxiv.org/abs/https://doi.org/10.1063/PT.3.3589}
  {https://doi.org/10.1063/PT.3.3589} \BibitemShut {NoStop}%
\bibitem [{\citenamefont {Zwickl}\ \emph {et~al.}(2013)\citenamefont {Zwickl},
  \citenamefont {Finkelstein},\ and\ \citenamefont {Lewandowski}}]{Zwickl13}%
  \BibitemOpen
  \bibfield  {author} {\bibinfo {author} {\bibfnamefont {B.~M.}\ \bibnamefont
  {Zwickl}}, \bibinfo {author} {\bibfnamefont {N.}~\bibnamefont
  {Finkelstein}},\ and\ \bibinfo {author} {\bibfnamefont {H.~J.}\ \bibnamefont
  {Lewandowski}},\ }\bibfield  {title} {\bibinfo {title} {The process of
  transforming an advanced lab course: Goals, curriculum, and assessments},\
  }\href {https://doi.org/10.1119/1.4768890} {\bibfield  {journal} {\bibinfo
  {journal} {American Journal of Physics}\ }\textbf {\bibinfo {volume} {81}},\
  \bibinfo {pages} {63} (\bibinfo {year} {2013})},\ \Eprint
  {https://arxiv.org/abs/https://doi.org/10.1119/1.4768890}
  {https://doi.org/10.1119/1.4768890} \BibitemShut {NoStop}%
\bibitem [{\citenamefont {Wilcox}\ and\ \citenamefont
  {Lewandowski}(2016)}]{Wilcox16}%
  \BibitemOpen
  \bibfield  {author} {\bibinfo {author} {\bibfnamefont {B.~R.}\ \bibnamefont
  {Wilcox}}\ and\ \bibinfo {author} {\bibfnamefont {H.~J.}\ \bibnamefont
  {Lewandowski}},\ }\bibfield  {title} {\bibinfo {title} {Open-ended versus
  guided laboratory activities:impact on students' beliefs about experimental
  physics},\ }\href {https://doi.org/10.1103/PhysRevPhysEducRes.12.020132}
  {\bibfield  {journal} {\bibinfo  {journal} {Phys. Rev. Phys. Educ. Res.}\
  }\textbf {\bibinfo {volume} {12}},\ \bibinfo {pages} {020132} (\bibinfo
  {year} {2016})}\BibitemShut {NoStop}%
\bibitem [{\citenamefont {Dounas-Frazer}\ \emph {et~al.}(2017)\citenamefont
  {Dounas-Frazer}, \citenamefont {Stanley},\ and\ \citenamefont
  {Lewandowski}}]{Dounas17}%
  \BibitemOpen
  \bibfield  {author} {\bibinfo {author} {\bibfnamefont {D.~R.}\ \bibnamefont
  {Dounas-Frazer}}, \bibinfo {author} {\bibfnamefont {J.~T.}\ \bibnamefont
  {Stanley}},\ and\ \bibinfo {author} {\bibfnamefont {H.~J.}\ \bibnamefont
  {Lewandowski}},\ }\bibfield  {title} {\bibinfo {title} {Student ownership of
  projects in an upper-division optics laboratory course: A multiple case study
  of successful experiences},\ }\href
  {https://doi.org/10.1103/PhysRevPhysEducRes.13.020136} {\bibfield  {journal}
  {\bibinfo  {journal} {Phys. Rev. Phys. Educ. Res.}\ }\textbf {\bibinfo
  {volume} {13}},\ \bibinfo {pages} {020136} (\bibinfo {year}
  {2017})}\BibitemShut {NoStop}%
\bibitem [{\citenamefont {Gosser}\ \emph {et~al.}(2001)\citenamefont {Gosser},
  \citenamefont {Cracolice}, \citenamefont {Kampmeier}, \citenamefont {Roth},
  \citenamefont {Strozak},\ and\ \citenamefont {Varma-Nelson}}]{Gosser}%
  \BibitemOpen
  \bibfield  {author} {\bibinfo {author} {\bibfnamefont {D.~K.}\ \bibnamefont
  {Gosser}}, \bibinfo {author} {\bibfnamefont {M.~S.}\ \bibnamefont
  {Cracolice}}, \bibinfo {author} {\bibfnamefont {J.~A.}\ \bibnamefont
  {Kampmeier}}, \bibinfo {author} {\bibfnamefont {V.}~\bibnamefont {Roth}},
  \bibinfo {author} {\bibfnamefont {V.~S.}\ \bibnamefont {Strozak}},\ and\
  \bibinfo {author} {\bibfnamefont {P.}~\bibnamefont {Varma-Nelson}},\
  }\href@noop {} {\emph {\bibinfo {title} {Peer-Led Team Learning: A
  Guidebook}}}\ (\bibinfo  {publisher} {Pearson},\ \bibinfo {year}
  {2001})\BibitemShut {NoStop}%
\bibitem [{\citenamefont {Walton}\ and\ \citenamefont
  {Cohen}(2011)}]{Walton11}%
  \BibitemOpen
  \bibfield  {author} {\bibinfo {author} {\bibfnamefont {G.~M.}\ \bibnamefont
  {Walton}}\ and\ \bibinfo {author} {\bibfnamefont {G.~L.}\ \bibnamefont
  {Cohen}},\ }\bibfield  {title} {\bibinfo {title} {A brief social-belonging
  intervention improves academic and health outcomes of minority students},\
  }\href {https://doi.org/10.1126/science.1198364} {\bibfield  {journal}
  {\bibinfo  {journal} {Science}\ }\textbf {\bibinfo {volume} {331}},\ \bibinfo
  {pages} {1447} (\bibinfo {year} {2011})},\ \Eprint
  {https://arxiv.org/abs/https://science.sciencemag.org/content/331/6023/1447.full.pdf}
  {https://science.sciencemag.org/content/331/6023/1447.full.pdf} \BibitemShut
  {NoStop}%
\bibitem [{\citenamefont {{Vision for Success Diversity, Equity and Inclusion
  Task Force}}()}]{CCCreport}%
  \BibitemOpen
  \bibfield  {author} {\bibinfo {author} {\bibnamefont {{Vision for Success
  Diversity, Equity and Inclusion Task Force}}},\ }\href@noop {} {\bibinfo
  {title} {2020 summary report}},\ \bibinfo {howpublished}
  {\url{https://www.cccco.edu/-/media/CCCCO-Website/Files/Communications/vision-for-success/cccco-dei-report.pdf}}\BibitemShut
  {NoStop}%
\bibitem [{NCH()}]{NCHEA}%
  \BibitemOpen
  \href@noop {} {\bibinfo {title} {North couty higher education alliance}},\
  \bibinfo {howpublished} {\url{https://www.ncheasd.com}}\BibitemShut {NoStop}%
\bibitem [{\citenamefont {{National Academies of Sciences, Engineering, and
  Medicine}}(2019{\natexlab{b}})}]{NAP25568}%
  \BibitemOpen
  \bibfield  {author} {\bibinfo {author} {\bibnamefont {{National Academies of
  Sciences, Engineering, and Medicine}}},\ }\href
  {https://doi.org/10.17226/25568} {\emph {\bibinfo {title} {The Science of
  Effective Mentorship in STEMM}}},\ edited by\ \bibinfo {editor}
  {\bibfnamefont {A.}~\bibnamefont {Byars-Winston}}\ and\ \bibinfo {editor}
  {\bibfnamefont {M.~L.}\ \bibnamefont {Dahlberg}}\ (\bibinfo  {publisher} {The
  National Academies Press},\ \bibinfo {address} {Washington, DC},\ \bibinfo
  {year} {2019})\BibitemShut {NoStop}%
\bibitem [{\citenamefont {Woolston}(2019)}]{Woolston19}%
  \BibitemOpen
  \bibfield  {author} {\bibinfo {author} {\bibfnamefont {C.}~\bibnamefont
  {Woolston}},\ }\bibfield  {title} {\bibinfo {title} {How four winning mentors
  help to build skills and dispel doubt},\ }\href@noop {} {\bibfield  {journal}
  {\bibinfo  {journal} {Nature}\ }\textbf {\bibinfo {volume} {565}},\ \bibinfo
  {pages} {255} (\bibinfo {year} {2019})}\BibitemShut {NoStop}%
\bibitem [{\citenamefont {Martonosi}\ and\ \citenamefont
  {Marrongelle}(2020)}]{DCL20101}%
  \BibitemOpen
  \bibfield  {author} {\bibinfo {author} {\bibfnamefont {M.}~\bibnamefont
  {Martonosi}}\ and\ \bibinfo {author} {\bibfnamefont {K.}~\bibnamefont
  {Marrongelle}},\ }\href@noop {} {\bibinfo {title} {Dear colleague letter:
  Advancing educational innovations that motivate and prepare prek-12 learners
  for computationally-intensive industries of the future}},\ \bibinfo
  {howpublished} {\url{https://www.nsf.gov/pubs/2020/nsf20101/nsf20101.jsp}}
  (\bibinfo {year} {2020})\BibitemShut {NoStop}%
\end{thebibliography}%

\end{document}